\documentclass[fleqn,usenatbib]{mnras}

\usepackage[T1]{fontenc}

\DeclareRobustCommand{\VAN}[3]{#2}
\let\VANthebibliography\thebibliography
\def\thebibliography{\DeclareRobustCommand{\VAN}[3]{##3}\VANthebibliography}

\usepackage{graphicx}	
\usepackage{amsmath}	
\usepackage{amssymb}	
\usepackage{soul}
\usepackage{txfonts}

\title[Branched carbon-chain molecules in hot core]{Astrochemical model to study the abundances of branched carbon-chain molecules in a hot molecular core with realistic binding energies}

\author[Srivastav et al.]{
Satyam Srivastav,$^{1}$
Milan Sil,$^{2}$
Prasanta Gorai,$^{3}$
Amit Pathak,$^{1}$
Bhalamurugan Sivaraman$^{4}$
\newauthor and Ankan Das$^{5}$\thanks{E-mail: ankan.das@gmail.com}
\\
$^{1}$ Department of Physics, Institute Of Science, Banaras Hindu University, Varanasi, 221005, India \\
$^{2}$ Department of Astrophysics and High Energy Physics, S. N. Bose National Centre for Basic Sciences, Block-JD, Sector-III, Salt Lake, Kolkata 700106, India \\
$^{3}$ Department of Space, Earth \& Environment, Chalmers University of Technology, SE-412 96 Gothenburg, Sweden \\
$^{4}$ Physical Research Laboratory, Navrangpura, Ahmedabad 380009, India \\
$^{5}$ Institute of Astronomy Space and Earth Science, AJ 316, Salt Lake, Sector II, Kolkata 700091, India}

\date{Accepted XXX. Received YYY; in original form ZZZ}

\pubyear{2022}

\begin{document}
\label{firstpage}
\pagerange{\pageref{firstpage}--\pageref{lastpage}}
\maketitle

\begin{abstract}
Straight-chain (normal-propyl cyanide, $\rm{n-C_3H_7CN}$) and branched-chain (iso-propyl cyanide, $\rm{i-C_3H_7CN}$) alkyl cyanides are recently identified in the massive star-forming regions (Sgr B2(N) and Orion). These branched-chain molecules indicate that the key amino acids (side-chain structures) may also be present in a similar region. The process by which this branching could propagate towards the higher-order (butyl cyanide, $\rm{C_4H_9CN}$) is an active field of research. Since the grain catalysis process could have formed a major portion of these species, considering a realistic set of binding energies are indeed essential. We employ quantum chemical calculations to estimate the binding energy of these species considering water as a substrate because 
water is the principal constituent of this interstellar ice. We find significantly lower binding energy values for these species than were previously used. It is noticed that the use of realistic binding energy values can significantly change the abundance of these species. The branching is more favorable for the higher-order alkyl cyanides with the new binding energies. With the inclusion of our new binding energy values and one essential destruction reaction ($\rm{i-C_3H_7CN+H \rightarrow CH_3C(CH_3)CN + H_2}$, having an activation barrier of $947$ K), abundances of $\rm{t-C_4H_9CN}$ dramatically increased.
\end{abstract}




\section{Introduction}
\label{sec:sample1}
About 270 molecules comprising 17 different elements have been detected in the interstellar medium (ISM) and circumstellar shells\footnote{\url{https://cdms.astro.uni-koeln.de/classic/molecules}}. H$_2$ is the most abundant molecule in the ISM, whereas CO is second \citep{wils70}. 
The protonated form of H$_2$, H$_3^+$ is also very plentiful. Despite this, 
numerous positively (e.g., HCO$^+$, CF$^+$) and negatively (e.g., CN$^-$, $\rm{C_3N^-}$, $\rm{C_4H^-}$) charged radicals, isomers (e.g., HNC, HCN), and isotopes (e.g., deuterium, $^{13}$C, $^{18}$O, $^{29}$Si) were also found.  The condensed regions of molecular clouds carry grains, which uphold some simple and complex molecules like H$_2$O, CO, CO$_2$, CH$_3$OH, H$_2$CO, NH$_3$, HCOOH, CH$_4$, etc. \citep{boog15,gora20a}. In the slightly warmer region, radicals become mobile and could produce various complex organic molecules \citep[COMs;][]{das08,das10,das11,das16}.
The major portion of the observed complex molecules in the ISM is organic, showing the active participation of the C atom in large molecules detected in space. 
The presence of larger PAHs ($\rm{C_nH_m}$) \citep{tiel08} and fullerenes (C$_{60}$, C$_{70}$) \citep{cami10} in space was obtained. 

The presence of complex organics indicates the existence of amino acids in the ISM \citep{sil18,gora20b}. Amino acids (building blocks of protein) are found in carbon chondrites and comets. For example, a carbonaceous meteorite, Murchison, provides more than 70 extraterrestrial amino acids \citep{bott02}. Detected organic molecules carry the aliphatic nature in a simple straight-chain structure (with less than three C atoms). In contrast, a branched carbon-chain structure also exists with four or more C atoms.

Several complex species of the prebiotic interest were recently been identified. Some of them are benzonitrile \citep{mcgu18}, propargylimine \citep{bizz20}, ethanolamine \citep{rivi21}, glycolonitrile \citep{zeng19}, ethynyl cyclopropenylidene, cyclopentadiene \citep{cern21a}, indene \citep{burk21,cern21a}, ethynyl cyclopentadiene \citep{cern21b}  1/2 - cyanocylopentadiene \citep{lee21,mcca21}, 1/2 - cyanonaphthalene \citep{mcgu21}, propylene oxide \citep[chiral molecule]{mcgu16}.

Iso-butyronitrile or iso-propyl cyanide ($\rm{i-C_3H_7CN}$) is 
is the first branched carbon-chain molecule (BCM), which has been observed in the high-mass star-forming regions (HMSFRs), Sagittarius B2 \citep[Sgr B2;][]{bell14}.
\cite{paga17} also detected $\rm{n/i-C_3H_7CN}$ for the first time in Orion.
The first detection of this kind of molecule has increased the curiosity of other BCMs in the HMSFRs. 
\cite{bell14} differentiated the mechanism of the formation of both conformers of $\rm{C_3H_7CN}$ and proposed their reaction pathways. They suggested the formation of $\rm{i-C_3H_7CN}$ via the addition of cyanide radical ($\rm{-CN}$) at the secondary carbon in the chain.

\cite{etim17} considered various isomers from the C$_5$H$_9$N isomeric group. Butyl cyanide ($\rm{C_4H_9CN}$) belongs to this isomeric group having four forms: the straight-chain normal-butyl cyanide ($\rm{n-C_4H_9CN}$), iso-butyl cyanide ($\rm{i-C_4H_9CN}$), sec-butyl cyanide ($\rm{s-C_4H_9CN}$), and tert-butyl cyanide ($\rm{t-C_4H_9CN}$). According to the expected rotational intensity ratio and considering equal abundances of these species, \cite{etim17} found that $\rm{t-C_4H_9CN}$ is the most favourable candidate among the $\rm{C_5H_9N}$ isomeric group for the future astronomical detection. However, interstellar chemistry is far from equilibrium, and reaction pathways for the formation of these species are very different. \cite{garr17} performed astrochemical modeling to decode how gas and grain are responsible for producing these species. They considered
a transformation of species from the surface to the mantle and vice versa.
They found $\rm{s-C_4H_9CN}$ with high abundance while $\rm{t-C_4H_9CN}$ with lesser abundance and proposed $\rm{s-C_4H_9CN}$ as future detectable BCM.
They implemented some educated guesses of binding energies (BE) to estimate their abundances. Here, in this work, our primary motivation is to study the fate of these species with realistic BEs. 
Since in the denser region, $\sim 70\%$ of the grain mantle is covered with water, we consider water as a substrate for the computation of the BEs of these species. Computed BEs are then used in our astrochemical model to refrain the abundances of these species.

This paper is organized as follows. Firstly, in Section \ref{sec:comp}, computational details, methodology, and reaction pathways are presented. Then, results and discussions are presented in Section \ref{sec:results}. Finally, in Section \ref{sec:conclusions}, we conclude.

\section{Computational details and methodology} \label{sec:comp}
\subsection{Binding Energy}
Merely a hundred years ago, the vast existence of anything but atoms and obscuring tiny dust grains in the ISM was unimaginable. 
The existence of interstellar dust grains confirmed by \cite{trum30} directed to exploring the presence of organic species on dust grains.
Species on grain surfaces generally undergo four different mechanisms:
a) Accretion (adsorption) onto the surface,
b) desorption from the surface,
c) diffusion across the surface or on/within the ice-mantle, and d) reaction.
Ice mantles are further processed when it is exposed to various interstellar radiations.

The adsorption energy or BE is the surface energy due to electrostatic interaction. It is the energy with which different particles or surfaces incline to attach. For example, it creates a film between the surface of species (adsorbate) and dust grain (adsorbent).
 We calculate the BE of a species as follows:
\[BE = (E_{\rm surface} + E_{\rm species}) - E_{\rm ss},\]
where $E_{ss}$ is the optimized energy of a species placed on the grain surface by a weak van der Waals interaction. $E_{\rm surface}$ and $E_{\rm species}$ are the optimized energies of the substrate and target species, respectively. Quantum chemical calculations are used to estimate the optimized energies of adsorbate and adsorbent. Since water molecules dominate the dense part of the interstellar ices \citep{kean01,das10,das11}, we consider it as a substrate. 

All the quantum chemical calculations are performed using the Gaussian 09 suite of programs \citep{fris13}.
We employ the second-order M\o ller-Plesset (MP2) method to optimize the geometries with the aug-cc-pVDZ basis set.
\cite{das18} carried out some benchmarking calculations for 16 stable species by considering ZPVE and BSSE corrections. Their computed values better agree with the experiments when ZPVE and BSSE corrections were not included. Here also, we do not consider ZPVE and BSSE corrections into account. The fully optimized structure of the species is further verified by checking the real harmonic vibrational frequencies.

Table \ref{tab:be} reports the computed BE of some species related to the formation of BCMs. The ground-state spin multiplicity of these species is also noted for clarity. \cite{das18} estimated BE of $\sim 100$ interstellar species. They proposed a scaling factor of 1.416 and 1.188 for water monomer and tetramer structure. Here, we use water monomer as a binding substrate for all species noted in Table \ref{tab:be}. But, sometimes, the size of the water monomer is minimal compared to the target species acting as an adsorbate. It may lead to some misleading estimations of BEs. For the betterment, we further use water tetramer as a substrate to estimate BEs of some species reported here.
\cite{das18,das21} noted that computed BEs are dependent on the chosen adsorption site.
Here, for some key species, we calculate BEs at multiple binding sites and noted in Table \ref{tab:be}. For the modeling purpose, we use the average scaled value. The BE values used by \cite{garr17} are also noted for comparison.

\begin{table*}
\scriptsize
{\centering
\caption{Calculated binding energy and enthalpy of formation of some key species and comparison with the previous estimation.}
    \label{tab:be}
    \begin{tabular}{cccccccccc}
    \hline
Serial &   & Ground &  \multicolumn{5}{c}{Binding Energies (K)} &  \multicolumn{2}{c}{Enthalpy of formation} \\
No. &   & State Spin &  \multicolumn{5}{c}{ } & \multicolumn{2}{c}{$\rm{\Delta H_f}$ (298 K) (kcal/mole)}\\ 
 & Species & Multiplicity & \multicolumn{4}{c}{This Work (MP2-aug-cc-pVDZ)}  & \cite{garr17}& This Work & \cite{garr17}\\ 
 & &  & with water monomer & Scaled by 1.416  & with water tetramer & Scaled by 1.188 & & G4 composite method & \\
   \hline
  1 &      $\rm{CH_3CN}$ & Singlet & 2614 & 3702 & & & 4680 &+15.84&+17.70 \\
  2 &      $\rm{C_2H_2CN}$ & Doublet & 3263* (2826 / 3700) & 4620* (4001 / 5240) & & & 4187 & +103.49& +105.84\\
  3 &      $\rm{CH_2CHCN}$ & Singlet & 2788* (2990 / 2587) & 3948* (4234 / 3663) & 2980 & 3540 & 4637 & +43.31 & +42.95 \\
  4 &      $\rm{\dot CH_2CH_2CN}$ & Doublet & 2900 & 4107 & & & 5087 & +61.54 & +55.13 \\
  5 &      $\rm{CH_3 \dot CHCN}$ & Doublet & 2730 & 3865 & & & 5087 & +47.14 & +53.23 \\
  6 &      $\rm{C_2H_5CN}$ & Singlet & 2867 & 4059 & 4113 & 4886 & 5537 & +10.96 & +12.71 \\
  7 &      $\rm{\dot CH_2CH_2CH_2CN}$ & Doublet & 3070 & 4347 & & & 6787&+54.61&+56.38\\
  8 &      $\rm{CH_3\dot CHCH_2CN}$ & Doublet & 3000* (3171 / 2830) & 4249* (4491 / 4007) & & & 6787 & +51.69 & +54.48 \\
  9 &      $\rm{CH_3CH_2\dot CHCN}$ & Doublet & 2738 & 3877 & & & 6787 & +42.23&+54.48\\
 10 &      $\rm{n-C_3H_7CN}$ & Singlet & 2112* (1307 / 2918) & 2991* (1850 / 4132) & 4686 & 5567 & 7237 & +5.56&+7.46 \\
 11 &      $\rm{\dot CH_2CH(CH_3)CN}$ & Doublet & 2958 &4188 & & & 6787&+55.26&+54.36\\
 12 &      $\rm{\dot CH_3\dot C(CH_3)CN}$ & Doublet & 3290* (3253 / 3328) & 4659* (4607 / 4711) & & & 6787 & +37.35 & +52.46 \\
 13 &      $\rm{i-C_3H_7CN}$ & Singlet & 2316* (1743 / 2888) & 3279* (2469 / 4090) & 4184 & 4970 & 7237 & +5.18 & +5.44 \\
 14 &      $\rm{CH_3CH_2\dot CHCH_2CN}$ & Doublet & 3224* (3423 / 3024) & 4564* (4847 / 4281) & & & 8487&+46.74& --- \\
 15 &      $\rm{n-C_4H_9CN}$ & Singlet & 4464 & 6320 & 3694 & 4388 & 8937&+0.50&+2.65 \\
 16 &      $\rm{i-C_4H_9CN}$ & Singlet & 3123 & 4422 & 4336 & 5151 & 8937&$-0.49$&+0.58\\
 17 &      $\rm{s-C_4H_9CN}$ & Singlet & 1861 & 2635 & 4472 & 5313 & 8937 &$-0.13$ & $+0.58$ \\
 18 &      $\rm{t-C_4H_9CN}$ & Singlet & 3230 & 4574 & 4333 & 5148 & 8937&$-0.83$&$-0.79$\\
 19 &      $\rm{\dot CH_2CH_2CH_3}$ & Doublet & 1866 & 2643 & & & 5637&+18.65&+23.90\\
 20 &      $\rm{CH_3\dot CHCH_3}$ & Doublet & 1532* (974 / 2091) & 2170* (1379 / 2961) & & & 5637&+14.56&+22.00\\
 21 &      $\rm{C_3H_8}$ & Singlet & 1028 & 1456 & & & 6087&$-29.93$&$-25.02$\\
 22 &      $\rm{\dot CH_2CH_2CH_2CH_3}$ & Doublet & 1464* (1883 / 1045) & 2073* (2666 / 1480) & & & 7337 & +13.82 & +17.90 \\
 23 &      $\rm{CH_3\dot CHCH_2CH_3}$ & Doublet & 2315 & 3278 & & & 7337 & +9.66&+16.00\\
 24 &      $\rm{n-C_4H_{10}}$ & Singlet & 1113 & 1576 & & & 7787 & $-35.05$ & $-30.03$ \\
 25 &      $\rm{\dot CH_2CH(CH_3)CH_3}$ & Doublet & 1057 & 1496 & & & 7337&+12.96&+17.00\\
 26 &      $\rm{CH_3\dot C(CH_3)CH_3}$ & Doublet & 2370 & 3357 & & & 7337 & +5.44 & +15.10 \\
 27 &      $\rm{i-C_4H_{10}}$ & Singlet & 1103* (1111 / 1095) & 1562* (1573 / 1551) & & & 7787 & $-35.90$ & $-32.07$ \\
 28 &      $\rm{n-C_5H_{12}}$ & Singlet & 3957 &5604 & & & 9487&$-40.05$&$-35.08$\\
 29 &      $\rm{i-C_5H_{12}}$ & Singlet & 992* (731 / 1254) & 1405* (1035 / 1775) & & & 9487 & $-40.05$ & $-36.73$ \\
 30 &      $\rm{neo-C_5H_{12}}$ & Singlet & 1028 & 1455 & & & 9487 & $-41.83$ &$-40.14$ \\
        \hline
    \end{tabular}} \\
*average of the BE values obtained from multiple binding sites. 
\end{table*}

\subsection{Other chemical parameters}
The enthalpies of formation are calculated at 298 K. The calculated enthalpies of formation are subsequently noted in Table \ref{tab:be} and compared with those indicated in \cite{garr17}. They noted these values from the NIST WebBook database and estimated where not available in the literature. The polarizability and dipole moment of these species are also calculated and compared (if available at the NIST WebBook database) in Table \ref{tab:polar_dipole}. These polarizabilities and dipole moments are further used in our model to obtain the destruction of these species by ion-neutral reactions \citep{su82,woon09}.

\subsection{Reaction rates}
Here, we prepare a reaction network to study the abundance of some BCMs. Most of the reaction rates of the formation BCMs are taken from \cite{garr17}. Here, we carry out transition state (TS) calculations with the Density Functional Theory (DFT) for some specific reactions. The ice-phase geometries of products, reactants, and TSs are optimized with the QST2 method and DFT-B3LYP/6-31+G(d) level of theory.
Similarly, the gas-phase geometries of products, reactants, and TSs are optimized using the Berny algorithm and DFT-B3LYP/6-31+G(d,p) and 6-31++G(2d,p) level of theories.
All TSs have a single imaginary frequency. The legitimacy of each calculated TS is verified by visually examining the vibrational mode corresponding to the single imaginary vibrational frequency and applying the criterion that it correctly connects the reactants and products through intrinsic reaction coordinate (IRC) paths. Finally, energy barriers are calculated using the TS theory. Formation pathways of the target molecules (vinyl, ethyl, i/n-propyl, i/n/s/t-butyl cyanide) are discussed in Section \ref{sec:EV}-\ref{sec:BC}.

\subsubsection{Vinyl and ethyl cyanide \label{sec:EV}}
Around the low-temperature regime, ice-phase formation of vinyl cyanide ($\rm{CH_2CHCN}$)  could process by the successive hydrogen additions of $\rm{HC_3N}$ (see reactions 1 and 2 in Table \ref{tab:EV}).
The first step of this H-addition reaction has an activation barrier of 1710 K (KIDA\footnote{\url{http://kida.astrophy.u-bordeaux.fr}}), whereas the second step is barrierless.
In the little warmer region, $\rm{CH_2CHCN}$ could also produce in the ice phase by the radical-radical reaction between CN and CHCH$_2$ (reaction 3).
The gas-phase reaction between CN and $\rm{C_2H_4}$ radicals (reaction 4) could contribute to the formation of $\rm{CH_2CHCN}$ beyond 100 K.
In the UMIST database \citep{mcel13}, this reaction is allowed only beyond 300 K with the three constants of the reaction: $\alpha=1.25 \times 10^{-10}$, $\beta=0.7$, and $\gamma=30.0$. However, the KIDA databse \citep{wake12} consider this reaction at the low temperature as well. For the lower limit of the temperature 10 K, they noted $\alpha=2.67 \times 10^{-10}$, whereas for 50 K, it is $5.31 \times 10^{-11}$ along with $\beta=-0.69$, and $\gamma= 31.0$.

\begin{table}
\scriptsize
{\centering
\caption{Ice-phase reactions considered for $\rm{CH_2CHCN}$ and $\rm{C_2H_5CN}$.}
  \label{tab:EV}
  \begin{tabular}{ccc}
    \hline
Reaction & Reactions & Activation Barrier (K) \\
Number (Type) &  &  \\
   \hline
\multicolumn{3}{c}{$\rm{CH_2CHCN}$} \\
\hline
1 (NR) & $\rm{H+ HC_3N \rightarrow C_3H_2N}$ & $1710^a$ \\
2 (RR) & $\rm{H+ C_3H_2N \rightarrow CH_2CHCN}$ & --- \\
3 (RR) & $\rm{CHCH_2 + CN \rightarrow CH_2CHCN}$ & --- \\
4 (RR) & $\rm{C_2H_4+CN \rightarrow CH_2CHCN+H}$ & --- \\
\hline
\multicolumn{3}{c}{$\rm{C_2H_5CN}$} \\
\hline
5 (NR) & $\rm{H + CH_2CHCN \rightarrow CH_3CHCN}$ & $619^b / 158^c$ \\
6 (NR) & $\rm{H + CH_2CHCN \rightarrow CH_2CH_2CN}$ & $1320^b / 1603^c$ \\
7 (RR) & $\rm{H+CH_3CHCN \rightarrow C_2H_5CN}$ & --- \\
8 (RR) & $\rm{H+CH_2CH_2CN \rightarrow C_2H_5CN}$ & --- \\
9 (RR) & $\rm{CH_3CH_2 +CN \rightarrow C_2H_5CN}$ & --- \\
10 (NR) & $\rm{C_2H_5CN+H \rightarrow C_2H_5CNH}$ & $2712^d$ \\
11 (NR) & $\rm{C_2H_5CN + H \rightarrow  CH_2CH_2CN + H_2}$ & $3472^c$ \\
\hline
\end{tabular}} \\
Notes: NR and RR refer to neutral-radical and barrierless radical-radical reactions, respectively. \\
$^a$ KIDA (\url{https://kida.astrochem-tools.org}) \\
$^b$ \cite{garr17} \\
$^c$ This work (gas-phase) \\
$^d$ \cite{sil18}
\end{table}

$\rm{CH_2CHCN}$ is further channelized to form ethyl cyanide ($\rm{C_2H_5CN}$) by successive hydrogenations in the ice phase (reactions $5-8$).
\cite{garr17} used an activation barrier of 619 K for the H-addition to the first carbon atom of $\rm{CH_2CHCN}$ (reaction 5) and 1320 K for the H-addition to the second carbon atom of $\rm{CH_2CHCN}$ (reaction 6).
\cite{garr17} adopted these activation barriers based on equivalent hydrogenation of $\rm{C_3H_6}$. Our TS calculations find an activation barrier of 158 K and 1603 K for reactions 5 and 6, respectively (see Fig. \ref{fig:reaction_5}). For our chemical model, we use our calculated values for reactions 5 and 6. At the same time, reactions 7 and 8 are considered barrierless.

In the warmer region, ice-phase formation of $\rm{C_2H_5CN}$ can follow the radical-radical reaction (reaction 9).
$\rm{C_2H_5CN}$ further hydrogenates to form ${\rm C_2H_5CNH}$ (reaction 10, with an activation barrier of 2712 K) which was considered in \cite{sil18}.

Furthermore, we explore one hydrogenation abstraction reaction of $\rm{C_2H_5CN}$ (reaction 11).
The reaction enthalpy of this reaction is $-3.38$ kcal/mol (DFT).
We obtain an activation barrier of 3472 K for this reaction in the gas phase.
However, the potential energy surface diagram of gas-phase reaction 11 shown in Fig. \ref{fig:ch2ch2cn} depicts that the energy of the products is less than that of reactants and TS does not converge in the ice phase. Due to these reasons, we do not include this reaction in our network.

\subsubsection{Propyl cyanide \label{sec:prop}}
Two isomeric forms of propyl cyanide ($\rm{n-C_3H_7CN}$ and $\rm{i-C_3H_7CN}$) are considered in our model. The formation pathways of these two isomeric forms are taken from \cite{garr17}. $\rm{n-C_3H_7CN}$ and $\rm{i-C_3H_7CN}$ formation in the ice phase could be processed by reactions $12-17$ and $18-21$, respectively, noted in Table \ref{tab:prop}.
For the destruction of ice-phase $\rm{i-C_3H_7CN}$, we consider a hydrogen abstraction reaction (reaction 22).
However, we could not converge the TS of reaction 22 in the ice phase, so here we use it in the gas phase.
The potential energy surface diagram of this reaction is shown in Fig. \ref{fig:ch3cch3cn}. We obtain an activation energy barrier of $947$ K for this abstraction reaction. The product of reaction 22 is again utilized in reaction 19 to form $\rm{i-C_3H_7CN}$ again. Reaction 22 is not considered as default in our network unless otherwise stated.  

\begin{table}
\scriptsize
{\centering
\caption{Ice-phase reactions considered for $\rm{C_3H_7CN}$ isomers.}
  \label{tab:prop}
  \begin{tabular}{ccc}
    \hline
Reaction & Reactions & Activation Barrier (K) \\
Number (Type) &  &  \\
   \hline
\multicolumn{3}{c}{$\rm{n-C_3H_7CN}$} \\
\hline
12 (RR) & $\rm{H  +       CH_2CH_2CH_2CN \rightarrow   n-C_3H_7CN}$ & --- \\
13 (RR) & $\rm{H   +      CH_3CHCH_2CN \rightarrow n-C_3H_7CN}$ & --- \\
14 (RR) & $\rm{H    +     CH_3CH_2CHCN \rightarrow n-C_3H_7CN}$ & --- \\
15 (RR) & $\rm{CH_3   +    CH_2CH_2CN  \rightarrow n-C_3H_7CN}$ & --- \\
16 (RR) & $\rm{CH_3CH_2   +   CH_2CN    \rightarrow  n-C_3H_7CN}$ & --- \\
17 (RR) & $\rm{CH_2CH_2CH_3 + CN       \rightarrow n-C_3H_7CN}$ & --- \\
\hline
\multicolumn{3}{c}{$\rm{i-C_3H_7CN}$} \\
\hline
18 (RR) & $\rm{H +        CH_2CH(CH_3)CN \rightarrow i-C_3H_7CN}$ & --- \\
19 (RR) & $\rm{H  +       CH_3C(CH_3)CN \rightarrow i-C_3H_7CN}$ & --- \\
20 (RR) & $\rm{CH_3 +      CH_3CHCN  \rightarrow i-C_3H_7CN}$ & --- \\
21 (RR) & $\rm{CN   +     CH_3CHCH_3  \rightarrow i-C_3H_7CN}$ & --- \\
22 (NR) & $\rm{i-C_3H_7CN + H \rightarrow  CH_3C(CH_3)CN + H_2}$ & $947^a$ \\
\hline
\end{tabular}} \\
Notes: NR and RR refer to neutral-radical and barrierless radical-radical reactions, respectively. \\
$^a$ This work (gas-phase).
\end{table}

\subsubsection{Butyl cyanide \label{sec:BC}}
Here, for the formation of the BCMs belonging to the $\rm{C_5H_9N}$ isomeric group, we consider the reaction pathways adopted in \cite{bell14,garr17}. They found that the radicals take a decisive part in their formation. These radicals were either produced by hydrogenations with carbon double bond or by hydrogen abstraction of a saturated carbon chain by the radicals like OH, NH$_2$, CH$_3$O, CH$_2$OH, etc. Normally, for the computation of the ice-phase reaction rates, the method proposed by \cite{hase92} is used. However, for the hydrogen abstraction reaction, this would underestimate the rate. To avoid this issue, in the context of hydrogen abstraction reactions, we use the mass of the hydrogen atom instead of the reduced mass of the reactants \citep{bell14,garr17}. Following \cite{gann07,garr17}, here also we consider low and high both the activation barriers for the reaction of $\rm{C_2H_2}$, ${\rm C_2H_4}$, $\rm{C_3H_6}$, and $\rm{C_4H_8}$ with the CN radical. Unless otherwise stated, we always use low barriers.

For the formation of four isomeric forms of $\rm{C_4H_9CN}$, our considered reactions are noted in Table \ref{tab:BC}. Interestingly, reaction 36 uses $\rm{CH_3C(CH_3)CN}$ as a reactant, which can be produced by our newly proposed hydrogen abstraction reaction of $\rm{i-C_3H_7CN}$ by reaction 22.

\begin{table}
\scriptsize
{\centering
\caption{Ice-phase reactions considered for $\rm{C_4H_9CN}$ isomers.}
  \label{tab:BC}
  \begin{tabular}{ccc}
    \hline
Reaction & Reactions & Activation Barrier (K) \\
Number (Type) &  &  \\
   \hline
\multicolumn{3}{c}{$\rm{n-C_4H_9CN}$} \\
\hline
23 (RR) & $\rm{CH_3     +  CH_2CH_2CH_2CN \rightarrow    n-C_4H_9CN}$ & --- \\
24 (RR) & $\rm{CH_2CH_3   +   CH_2CH_2CN  \rightarrow n-C_4H_9CN}$ & --- \\
25 (RR) & $\rm{CH_2CH_2CH_3    +  CH_2CN   \rightarrow  n-C_4H_9CN}$ & --- \\
26 (RR) & $\rm{CH_2CH_2CH_2CH_3  +     CN    \rightarrow    n-C_4H_9CN}$ & --- \\
27 (RR) & $\rm {H        + CH_3CH_2CHCH_2CN  \rightarrow  n-C_4H_9CN}$ & --- \\
\hline
\multicolumn{3}{c}{$\rm{i-C_4H_9CN}$} \\
\hline
28 (RR) & $\rm{CH_3   +    CH_3CHCH_2CN \rightarrow i-C_4H_9CN}$ & --- \\
29 (RR) & $\rm{CH_3CHCH_3 + CH_2CN  \rightarrow   i-C_4H_9CN}$ & --- \\
30 (RR) & $\rm{CN      +  CH_2CH(CH_3)CH_3 \rightarrow  i-C_4H_9CN}$ & --- \\
\hline
\multicolumn{3}{c}{$\rm{s-C_4H_9CN}$} \\
\hline
31 (RR) & $\rm{CH_3     +  CH_3CH_2CHCN   \rightarrow s-C_4H_9CN}$ & --- \\
32 (RR) & $\rm{CH_3      + CH_2CH(CH_3)CN \rightarrow s-C_4H_9CN}$ & --- \\
33 (RR) & $\rm{CH_2CH_3     + CH_3CHCN \rightarrow s-C_4H_9CN}$ & --- \\
34 (RR) & $\rm{CN       + CH_3CHCH_2CH_3 \rightarrow s-C_4H_9CN}$ & --- \\
\hline
\multicolumn{3}{c}{$\rm{t-C_4H_9CN}$} \\
\hline
35 (RR) & $\rm{CH_3C(CH_3)CH_3 +    CN    \rightarrow   t-C_4H_9CN}$ & --- \\
36 (RR) & $\rm{CH_3C(CH_3)CN   + CH_3   \rightarrow   t-C_4H_9CN}$ & --- \\
\hline
\end{tabular}} \\
Notes: RR refers to barrierless radical-radical reactions. \\
\end{table}

\subsection{Physical condition}
We consider a free-fall collapsing cloud followed by a warm-up and post-warm-up phases for our physical model \citep{garr13,gora20b,das21,sil21}. During the collapsing phase ($\rm{t_{coll}}$), total hydrogen density (n$_{\rm H}$) can evolve from a low density ($3 \times 10^3$ cm$^{-3}$) to a higher density ($2 \times 10^8$ cm$^{-3}$). The highest density attained at the collapsing phase is kept constant throughout the warm-up and post-warm-up phases. The choice of our highest density is consistent with the density derived from the $\rm{C_3H_7CN}$ emission from the core N2 of Sgr B2 \citep{bell14}. Following \cite{garr08}, we consider that the density and visual extinction parameter ($A_{\rm V}$) is coupled by $\rm{A_{\rm V}=A_{V0}(n_H/n_{H0})^{2/3}}$. Here, $\rm{A_{V0}}$ ($=2$) and n$_{\rm H0}$ ($=3 \times 10^3$ cm$^{-3}$) is the minimum visual extinction and total hydrogen density considered in our model. Using the highest density (n$_{\rm H}=2 \times 10^8$ cm$^{-3}$), in our case, A$_{\rm V}$ could reach a value as high as $3288$. The dust temperature (T$_{\rm dust}$) is derived by the relation provided by \cite{zucc01} and modified by \cite{garr08}. 
$$\rm{
T_{dust}=18.67-1.637 A_V+0.07518 {A_V}^2 -0.001492 {A_V}^3}.
$$
The above relation holds for A$_{\rm V}=2-10$. For A$_{\rm V}=2$, it yields $\rm{T_{dust}} \sim 16 $ K. The dust temperature further decreases as the visual extinction increases. Here, we restrict $\rm{T_{dust}}$ to fall below 8 K. At this phase, the gas temperature ($\rm{T_{gas}}$) is kept constant at 10 K. In the warm-up phase, $\rm{T_{dust}}$ is allowed to increase to 200 K in $\rm{t_{warm1}}$ years. Furthermore, to follow the further evolution, the dust temperature is allowed to increase up to 400 K in another $\rm{t_{warm2}}$ years. Once the dust temperature crosses the gas temperature, gas temperature follows the dust temperature because of the good coupling between the gas and dust at a higher density. Very similar warm-up time scales were considered in \cite{garr13,garr17}. In the post-warm-up phase (for $\rm{t_{pw}}$ years), all the physical parameters are kept constant at their respective highest values.

\begin{figure*}
\centering
\includegraphics[width=5 cm,angle=-90]{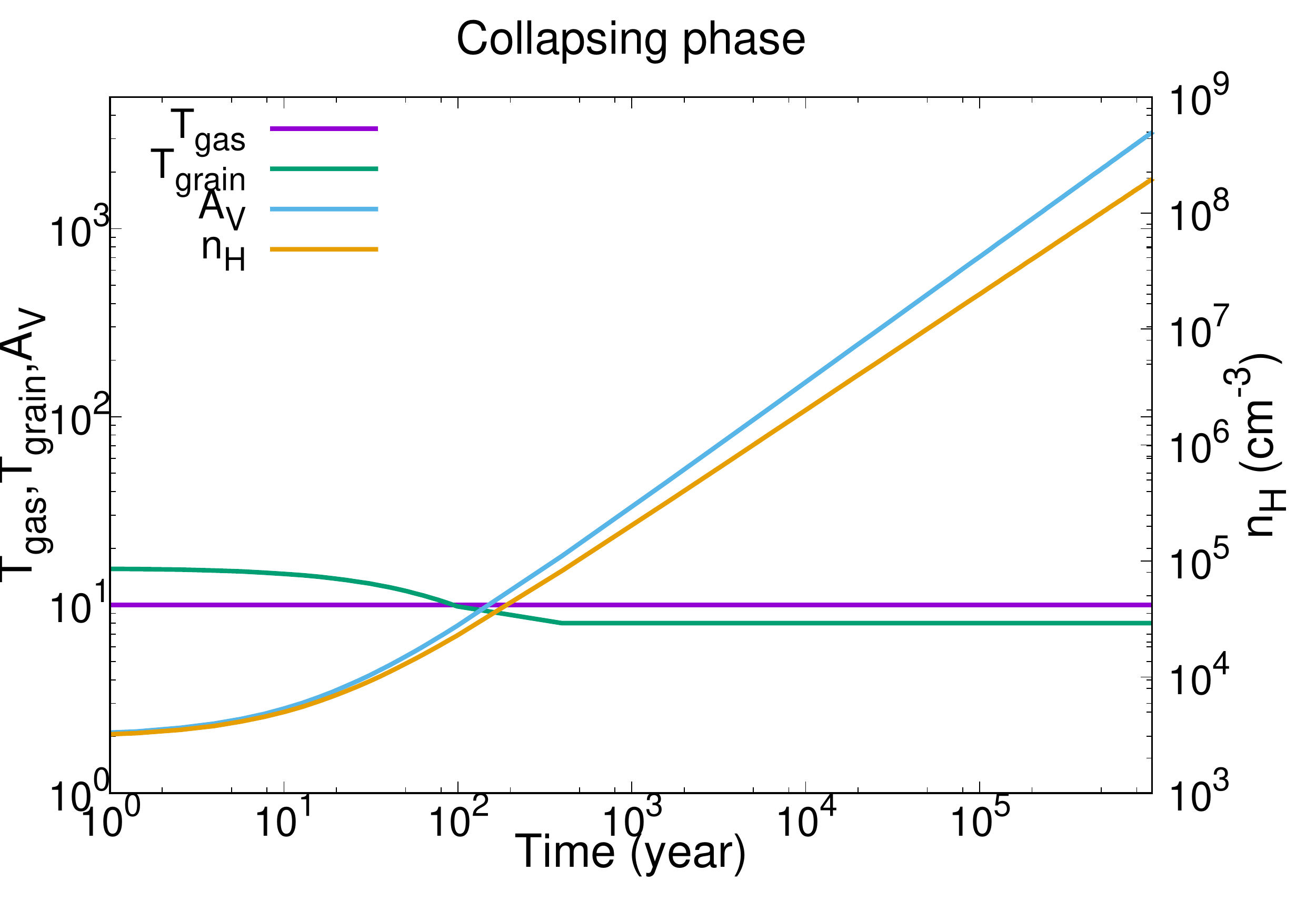}
\includegraphics[width=5 cm,angle=-90]{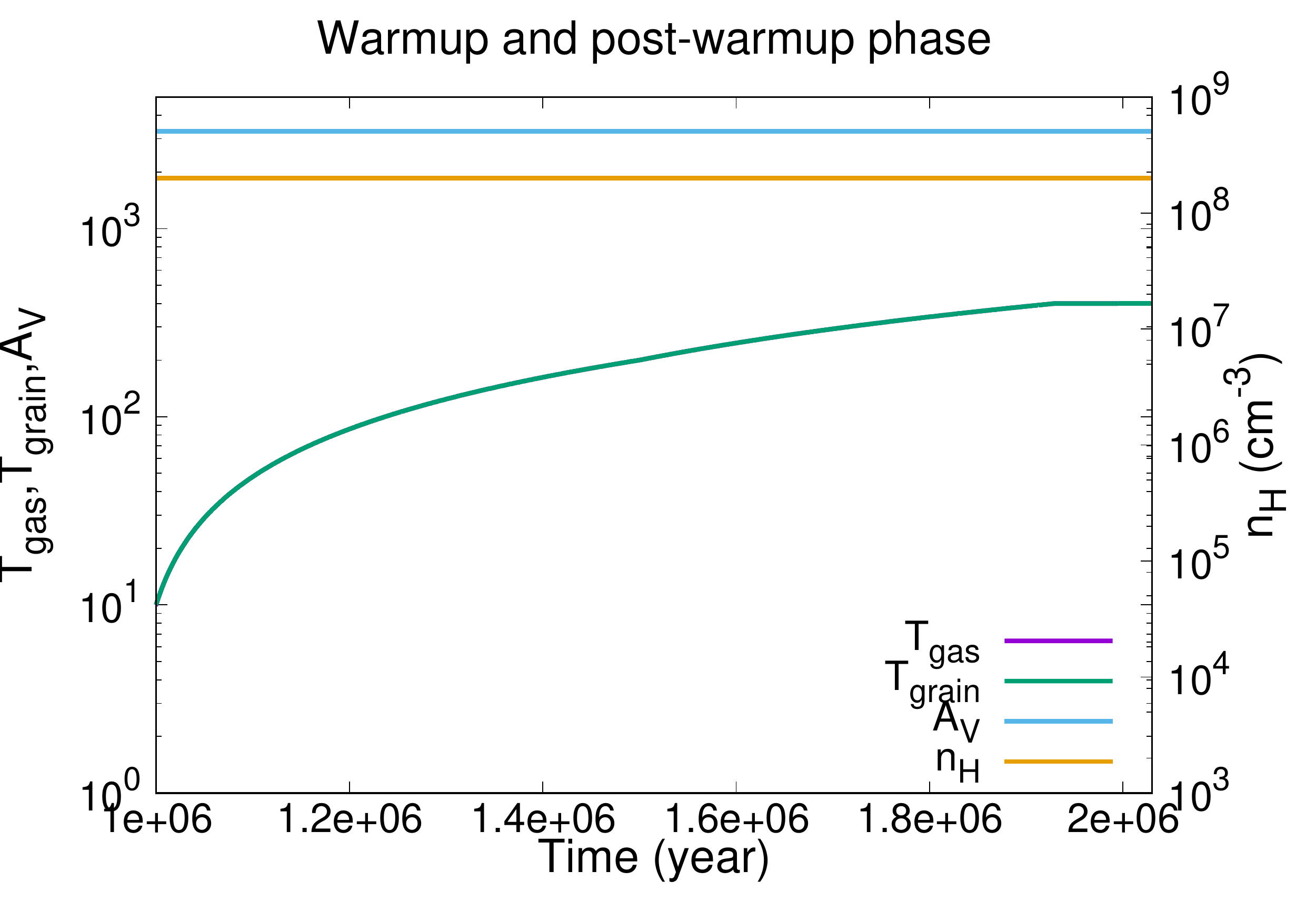}
\caption{The left panel shows the physical parameters ($\rm{T_{gas}}$, $\rm{T_{grain}}$, $\rm{A_V}$, and $\rm{n_H}$) for the isothermal phase, and the right panel shows the same at the warm-up and post-warm-up stages for Model A, where $\rm{t_{warm1}}=5 \times 10^5$ years.
\label{fig:phys}}
\end{figure*}

The collapsing and warm-up time would differ between the high mass and low-mass stars. A shorter collapsing time is expected for a high-mass star, whereas relatively longer for a low-mass star. Here, we construct models to explain the abundances observed in Sgr B2.
Based on these time scales, we consider two models (Model A and Model B) to explain the observed results.
The time scales considered in each model are shown in Table \ref{tab:models}. In Model A, the first warm-up time scale ($t_{warm1}$) is varied, whereas, in Model B, the collapsing time is varied by keeping all other time scales at the fixed value. Fig. \ref{fig:phys} represents the time evolution of all the physical parameters considered in this simulation for Model A only. For all the models, we consider a standard cosmic ray ionization rate of $1.3 \times 10^{-17}$ s$^{-1}$.

\begin{table}
\centering
 \caption{Adopted models based on various time scale.}
  \label{tab:models}
    \begin{tabular}{c|c|c}
    \hline
        Time &Model A& Model B  \\
         \hline
         $t_{coll}$& $10^6$ & $10^5-10^6$\\
         $t_{warm1}$& $10^5-10^6$ &$5 \times 10^5$\\
         $t_{warm2}$& $4.3 \times 10^5$ &$2.12 \times 10^4$\\
         $t_{pw}$& $10^5$&$10^5$\\
         \hline
         Total Time &$(1.63-2.53) \times 10^6$&$(0.72 -1.62) \times 10^6$\\
         \hline
             \end{tabular}
\end{table}

\section{Chemical model results and discussion}
\label{sec:results}
Here, we use our  CMMC (Chemical Model of Molecular Cloud) code \citep{das15a,das15b,das19,das21,gora17a,gora17b,gora20b,sil18,sil21,ghos22} to study the formation of BCMs in Sgr B2(N). Three sets of BE values are used. For all sets, we consider the ratio between the energy for diffusion and energy for desorption ($E_b/E_D$) at 0.5. The set 1 is constructed with the BE values used in \cite{garr17,bell14}. The enthalpies of formation of 
$\rm{CH_2CHCN}$, $\rm{C_2H_5CN}$, $\rm{C_3H_7CN}$, $\rm{C_4H_9CN}$, and their related precursors are also used from \cite{garr17,bell14}. Set 2 is constructed with the same BE values used in set 1 except those are reported in Table \ref{tab:be}. Whenever we use set 2 BE, we also use our calculated enthalpy of formation values reported in Table \ref{tab:be}.
Table \ref{tab:be} shows the BE of 30 relevant species with the monomer water configuration. However, for the eight species ($\rm{CH_2CHCN}$, $\rm{C_2H_5CN}$, $\rm{n/i-C_3H_7CN}$, $\rm{n/i/s/t-C_4H_9CN}$), we note the BE values with the tetramer configuration of water. 
Where monomer structure is used, we use a scaling factor of 1.416, and for the tetramer structure, a scale factor of 1.188 is used \citep{das18}.
Finally, in set 3, we keep the BEs and enthalpies of formation of these 30 species (noted in Table \ref{tab:be}) same as set 2, but for the rest of the species, we use the BEs from the KIDA database. Several differences exist between the BE used in set 2 and set 3. But the significant difference which could alter the abundances of the saturated species on the grain surface is the usage of the slow diffusion rate of the H atom in set 3 ($E_D=650$ K) compared to set 2 ($E_D=450$ K). 

\begin{figure*}
    \centering
    \includegraphics[width=11cm, angle= 270]{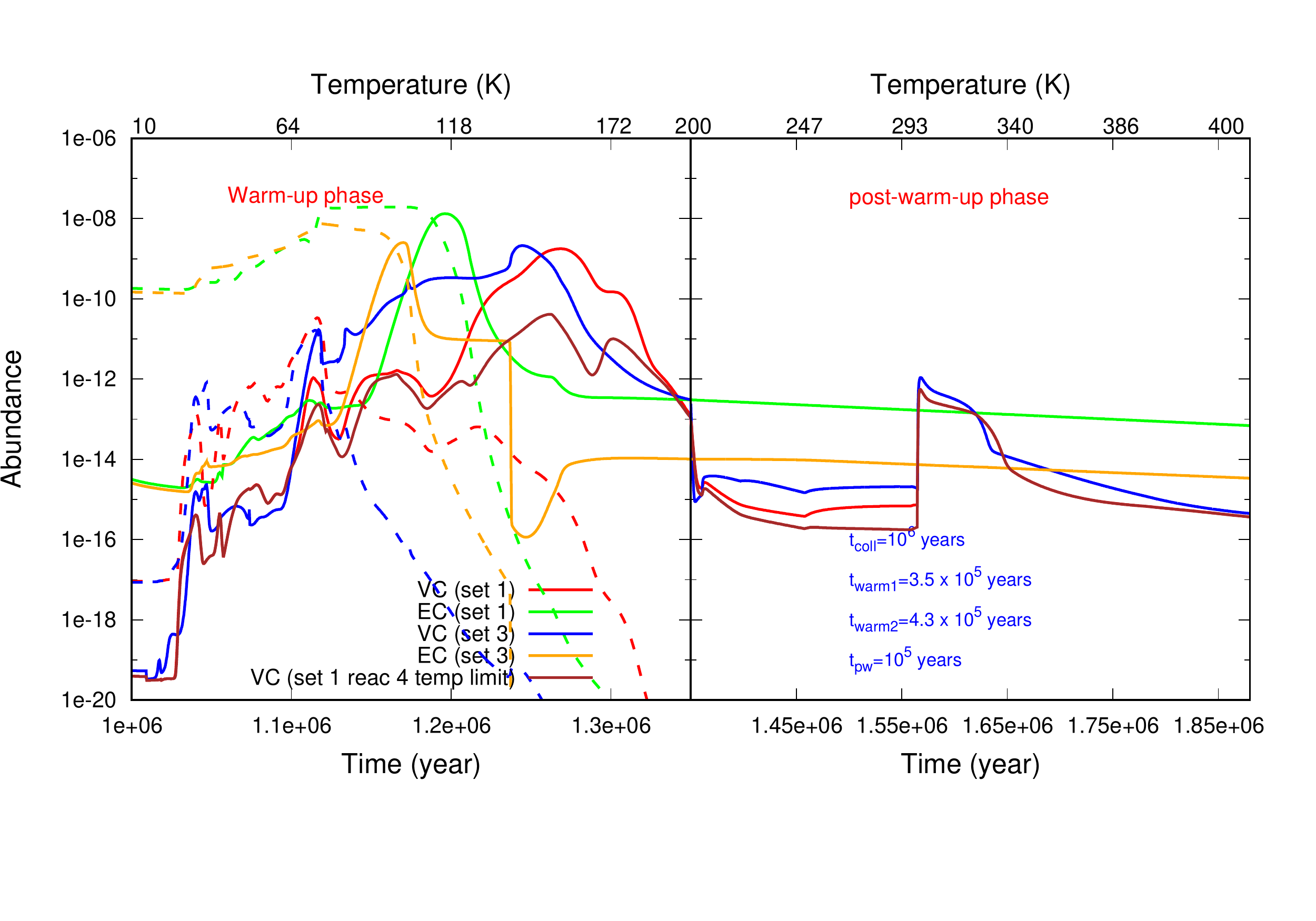}
    \caption{Time evolution of $\rm{CH_2CHCN}$ (VC) and $\rm{C_2H_5CN}$ (EC) during the warm-up and post-warm-up phase for set 1 and set 3 BEs. Solid curves represent the gas-phase abundances, whereas the dashed curves represent the abundances of ice-phase species. When the temperature limit is implemented, the solid brown curve represents the gas-phase abundance of $\rm{CH_2CHCN}$. It shows a decline in $\rm{CH_2CHCN}$.}
    \label{fig:EV}
\end{figure*}

\begin{figure*}
    \centering
    \includegraphics[width=5 cm,angle= 270]{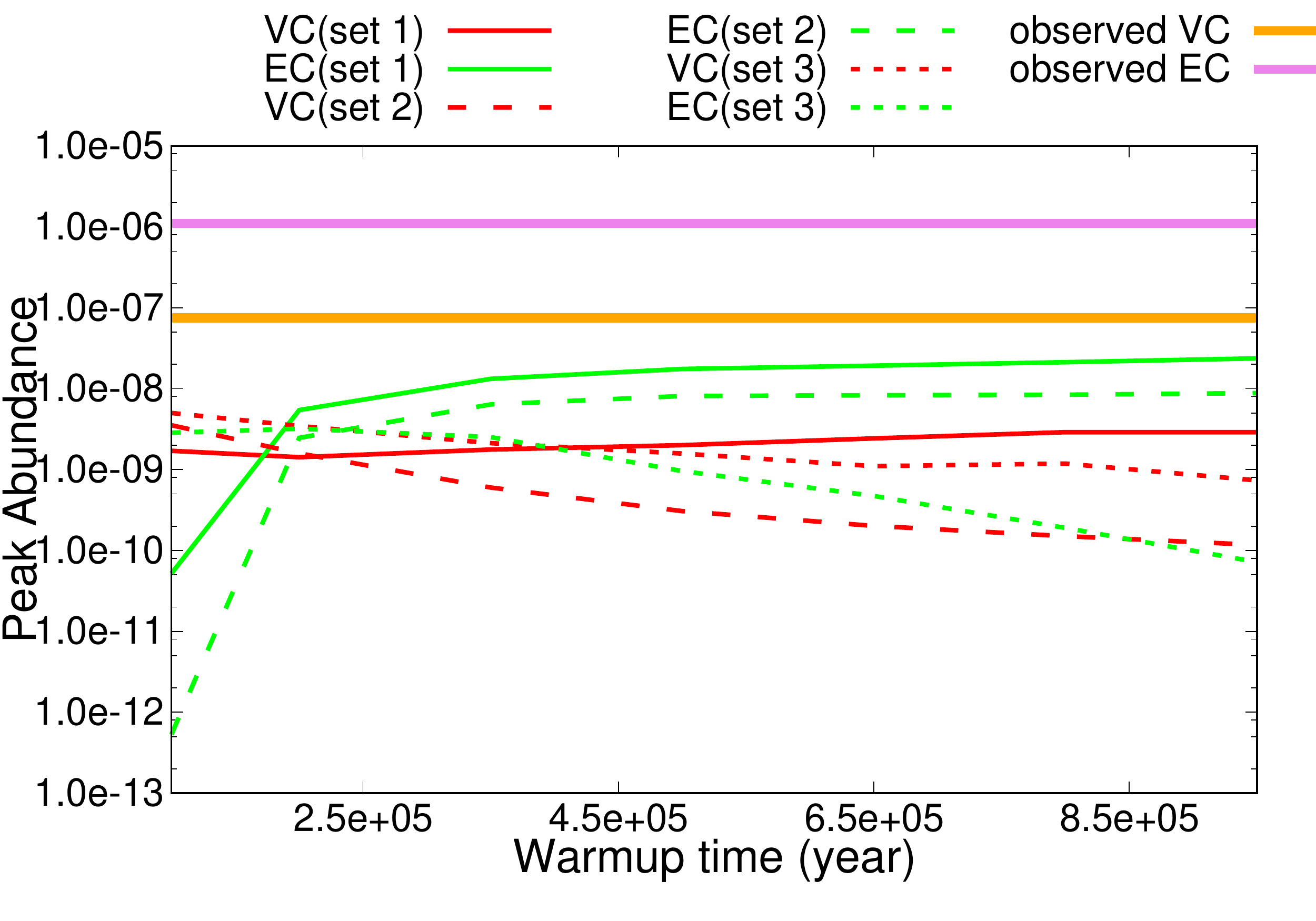}
    \includegraphics[width=5 cm,angle= 270]{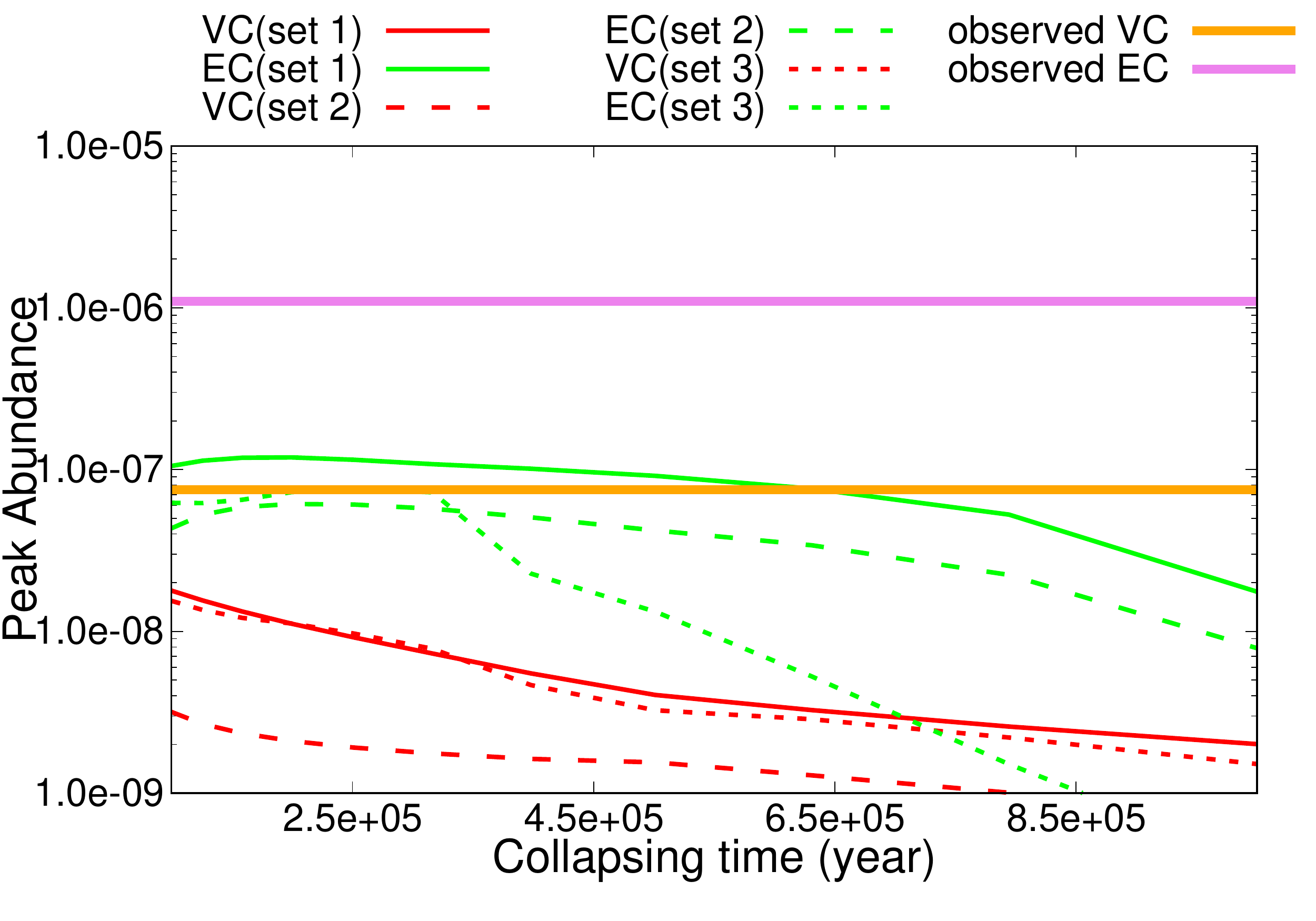}
    \caption{Peak abundances obtained beyond the collapsing phase for $\rm{CH_2CHCN}$ (VC) and $\rm{C_2H_5CN}$ (EC). The left panel shows results for Model A and right panel for Model B. Observed peak abundances of VC and EC in Sgr B2(N) are also shown with the horizontal solid lines.}
    \label{fig:EV-coll}
\end{figure*}

\begin{table*}
\caption{Peak abundances (with respect to H$_2$) and corresponding temperature (in K) obtained from our simulation for Model A with $\rm{t_{warm1}}=3.5 \times 10^5$ years.  \label{tab:pk-abn}}
   { \centering
    \begin{tabular}{|c|c|c|c|}
    \hline
                 &set 1&set 2&set 3\\
                 &&low/high&low/high/low(reaction 22)\\
    \hline
         $\rm{CH_2CHCN}$ & $1.8 \times 10^{-9}, 7.5 \times 10^{-8o}, 7.4 \times 10^{-9g}$ (156, 200$^o$, 167$^g$) & $6.0/6.3 \times 10^{-10} (126)$ & $2.12/2.20/2.06 \times 10^{-9} (143)$ \\
         $\rm{C_2H_5CN}$ & $1.3 \times 10^{-8}, 1.1 \times 10^{6o}, 3.5 \times 10^{-8g}$ (116, 150$^o$, 129$^g)$ & $6.4/6.7 \times 10^{-9} (100)$ & $2.52/2.46/3.18 \times 10^{-9} (102)$ \\
         $\rm{n-C_3H_7CN}$ & $3.8 \times 10^{-9}, 1.3 \times 10^{-8b}, 1.9 \times 10^{-9g}$ (146, 153$^b$, 156$^g)$ & $5.5/6.1 \times 10^{-10} (106)$ & $1.64/1.94/1.33 \times 10^{-9} (109)$ \\
         $\rm{i-C_3H_7CN}$ & $4.5 \times 10^{-9}, 3.2 \times 10^{-8b}, 3.4 \times 10^{-9g} (146, 153^b, 154^g)$ & $10.4/11.4 \times 10^{-10} (108)$ & ${2.01/2.34/1.61} \times 10^{-9}$ (110) \\
         $\rm{n-C_4H_9CN}$ & $2.3 \times 10^{-10}, 2.0 \times 10^{-9g} (175,190^g)$ & $1.88/1.94 \times 10^{-10} (103)$ & $7.01/6.96/8.2 \times 10^{-11} (105)$\\
         $\rm{i-C_4H_9CN}$ & $2.2 \times 10^{-10}, 3.8 \times 10^{-9g} (175,190^g)$ & $8.39/8.63 \times 10^{-11} (101)$ & $1.04/1.09/0.91 \times 10^{-9} (101)$\\
         $\rm{s-C_4H_9CN}$&$6.5 \times 10^{-10}, 3.7 \times 10^{-9g} (175,190^g)$ &$2.93/2.97 \times 10^{-10} (105)$ & $6.78/6.57/7.3 \times 10^{-10}(106)$ \\
         $\rm{t-C_4H_9CN}$& $1.3 \times 10^{-10}, 2.0 \times 10^{-10g} (175,190^g)$ & $1.81/1.53 \times 10^{-11} (101)$ & $7.37/1.22/129 \times 10^{-11} (101)$\\
         \hline
    \end{tabular}} \\
    $^g$ slow warm-up model with set 1 BEs and low activation \citep{garr17}\\
    $^o$\cite{bell16}\\
    $^b$\cite{bell14}\\
\end{table*}

\begin{table*}
{\scriptsize 
    \caption{Ranges of peak abundance ratio obtained from our Model A and Model B with various BEs and their comparison with literature. \label{tab:ratios}}
   \centering
    \begin{tabular}{|c|c|c|c|c|}
    \hline
         & Set 1 & Set 2 & Set 3 & Observed / other results  \\
         &low/high&low/high&low/high&\\
         \hline
       ${\rm C_2H_5CN/CH_2CHCN}$ &$5.9-23.4/0.033-22.5$&$13.7-32.5/0.00014-72$&$0.1-9.27/0.1-8.52$&15$^b$, $3.8-9.5^g$\\
       ${\rm i-C_3H_7CN/n-C_3H_7CN}$&${1.0-1.2/ 0.78-1.36}$&${1.8-3.5/1.09-3.4}$&${0.7-1.22/0.7-1.21}$&$0.4\pm 0.06^a$, $0.17-3.0^g$, $0.17-0.5^p$\\
       ${\rm n-C_3H_7CN/C_2H_5CN}$&${0.04-0.18/0.046-70.3}$&${0.005-0.059/0.0065-3399}$&${0.036-7.74/0.04-7.9}$&$0.029^{a,b}$, $0.024-0.67^g$\\
        ${\rm tot-C_3H_7CN/C_2H_5CN}$& ${0.078-0.37/0.086-166}$ &${0.023-0.17/0.0289-7114}$&${0.07-16.4/0.073-16.71}$&$0.041^{a,b}$, $0.98-0.21^g$\\
         ${\rm n-C_4H_9CN/n-C_3H_7CN}$&${0.033-0.052/0.03-0.3}$ &${0.34-1.0/0.015-0.77}$&${0.015-0.29/0.018-0.25}$&$<0.59$,$0.1-1.1^g$\\
         ${\rm i-C_4H_9CN/n-C_4H_9CN}$&${0.69-2.28/0.70-2.37}$&${0.12-0.44/0.16-14.3}$&${0.46-67/0.57-48.91}$&$0.6-2.2^g$\\
         ${\rm s-C_4H_9CN/n-C_4H_9CN}$&${2.45-3.0/2.23-3.03}$&${0.98-1.92/1.02-9.64}$&${3.36-24.8/3.46-17.73}$&$1.7-4.3^g$\\
         ${\rm t-C_4H_9CN/n-C_4H_9CN}$&${0.23-1.04/0.19-1.52}$&${0.014-0.12/0.014-5.34}$&${0.025-5.92/0.007-1.91}$&$0.015-0.1^g$\\
         ${\rm (i+s+t)-C_4H_9CN/n-C_4H_9CN}$&${3.45-6.38/3.2-5.9}$&${1.12-2.37/1.21-29.3}$&${3.84-97.64/4.04-66.7}$&$3.0-5.8^g$\\
        \hline
    \end{tabular}} \\
     $^a$\cite{bell14},
     $^b$\cite{bell16},
     $^g$\cite{garr17},
     $^p$\cite[Orion KL]{paga17},
     $^h$High activation
   \end{table*}

\subsection{Vinyl and ethyl cyanide}

Table \ref{tab:EV} shows the reactions leading to the formation of $\rm{CH_2CHCN}$ and $\rm{C_2H_5CN}$ in the interstellar condition.
Set 1 is constructed with the BE used in \cite{garr17,bell16}. They used higher BE of $\rm{C_2H_5CN}$ (5537 K) than $\rm{CH_2CHCN}$ (4637 K). We also obtain a higher BE of $\rm{C_2H_5CN}$ (4059 K) with the water monomer than $\rm{CH_2CHCN}$ (3948 K). However, the water monomer configuration is relatively smaller (3 atoms) than the adsorbed species (7 atoms for $\rm{CH_2CHCN}$ and 9 atoms for $\rm{C_2H_5CN}$), which could induce some errors on the estimated BEs.
\cite{das18} reported that the BE estimation tends to more realistic values as the size of the computed substrate is increased.
To check the effect of BEs on the larger substrate, we further use a tetramer water structure (consisting of 12 atoms). As like the monomer configuration, here also, we found that the BE of $\rm{C_2H_5CN}$ (scaled value $\sim 4886$ K) is greater than $\rm{CH_2CHCN}$ (scaled value $\sim 3540$ K).
We notice our estimated BE values with the tetramer configuration of water are lower by several hundreds of Kelvin than that of \cite{garr17} for both the species. 
Set 2 and 3 consider these BE values and enthalpy of formation noted in Table \ref{tab:be} for our simulation. 
 
Fig. \ref{fig:EV} shows the time evolution of $\rm{CH_2CHCN}$ and $\rm{C_2H_5CN}$ in the warm-up and post-warm-up phase for set 1 and set 3 BEs. The solid lines represent the gas-phase abundances, whereas the dashed lines represent the ice-phase abundances. Model A with $\rm{t_{warm1}}=3.5 \times 10^5$ years is used for Fig. \ref{fig:EV}. 
The gas-phase peak abundances (beyond the collapsing time) obtained with various sets of BEs are noted in Table \ref{tab:pk-abn}.
We obtain a higher peak abundance of $\rm{C_2H_5CN}$ than $\rm{CH_2CHCN}$ for all the cases. 
$\rm{C_2H_5CN}$ was mainly produced in the ice phase by the successive hydrogenations of $\rm{CH_2CHCN}$ (reactions $5-8$) and by the radical-radical reaction between $\rm{CH_3CH_2}$ and CN (reaction 9). With the set 1 BE, a peak abundance of $\rm{C_2H_5CN}$ is obtained at 116 K, whereas, with similar BEs, \cite{garr17} got the peak abundance at 129 K (they used $\rm{t_{warm1}= 10^6}$ years). 
For the set 2 and set 3, it varies in the range $100-102$ K for $\rm{t_{warm1}= 3.5 \times 10^5}$ years. 

The peak abundance of $\rm{CH_2CHCN}$ for set 1 is obtained at 156 K \citep[][got this at 167 K]{garr17}.
We notice a substantial effect of reaction 4 on $\rm{CH_2CHCN}$.
According to the UMIST Database for Astrochemistry 2012 \citep{mcel13}, this reaction is valid beyond 300 K. However, following KIDA database \citep{wake12}, we consider that this reaction may process beyond 10 K. The $\alpha$, $\beta$, and $\gamma$ noted in the UMIST database is used as the rate constants of this reaction. The gas-phase abundance profile of $\rm{CH_2CHCN}$ by utilizing the temperature restriction is shown in Fig. \ref{fig:EV} with the solid brown curve. It clearly shows that with the temperature restriction of this reaction, gas-phase peak $\rm{CH_2CHCN}$ abundance drops from $1.8 \times 10^{-9}$ (not considering the temperature limit) to $4.12 \times 10^{-11}$ (considering the temperature limit). The abundance of $\rm{CH_2CHCN}$ and $\rm{C_2H_5CN}$ noted in Table \ref{tab:pk-abn} shows a very minor change between the usage of low/high barrier for the CN addition to $\rm{C_2H_2}$, $\rm{C_2H_4}$, $\rm{C_3H_6}$, and $\rm{C_4H_8}$).

\cite{bell14} derived a peak H$_2$ column density of $4.2 \times 10^{24}$ cm$^{-2}$.
Furthermore, they extrapolated their obtained column density to a more compact
region ($\sim 100$) where the $\rm{C_3H_7CN}$ emission originated.
With this consideration, they estimated an average H$_2$ column density of $5.6 \times 10^{24}$ cm$^{-2}$. They identified $154$ transitions of $\rm{C_2H_5CN}$ and 44 transitions of $\rm{CH_2CHCN}$. They estimated a column density of $\rm{C_2H_5CN}$ and $\rm{CH_2CHCN}$ of $6.2 \times 10^{18}$ and $4.2 \times 10^{17}$, respectively.
Transforming into the abundances, it yields the abundances of $1.1 \times 10^{-6}$ and $7.5 \times 10^{-8}$ for $\rm{C_2H_5CN}$ and $\rm{CH_2CHCN}$, respectively.
Fig. \ref{fig:EV-coll} shows the variation of gas-phase peak abundances
(obtained beyond the collapsing time) of $\rm{CH_2CHCN}$ and $\rm{C_2H_5CN}$ with the changes in warm-up time scale (Model A, left panel) and collapsing time (Model B, right panel).
Interestingly, in most cases, the peak abundance of $\rm{C_2H_5CN}$ is greater
than that of the $\rm{CH_2CHCN}$ with Model A and Model B with various BEs.
With the set 1 BEs (red solid curve in the right panel of Fig. \ref{fig:EV-coll}),
the gas-phase peak abundance of $\rm{CH_2CHCN}$ and $\rm{C_2H_5CN}$ varies in
the range $1.42 \times 10^{-9} - 1.48 \times 10^{-8}$ and $5.19 \times 10^{-11} - 1.19 \times 10^{-7}$, respectively.
Only for the lowest warm-up time ($t_{warm1} = 10^5$ years) and set 1, we obtain VC>EC. None of our models (with set 2 and set 3 BEs) can
reproduce the observed abundance of $\rm{C_2H_5CN}$ ($1.1 \times 10^{-6}$) and
$\rm{CH_2CHCN}$ ($7.5 \times 10^{-8}$) in Sgr B2(N2). From the various models (fast, medium, and slow warm-up along with the low and high activation barriers), \cite{garr17} also obtained a comparatively lower abundance of $\rm{C_2H_5CN}$ ($1.2 \times 10^{-8} - 1.1 \times 10^{-7}$) and $\rm{CH_2CHCN}$ ($1.6 \times 10^{-9} - 1.7 \times 10^{-8}$) than observations.

\begin{figure*}
    \centering
    \includegraphics[width=5 cm, angle= 270]{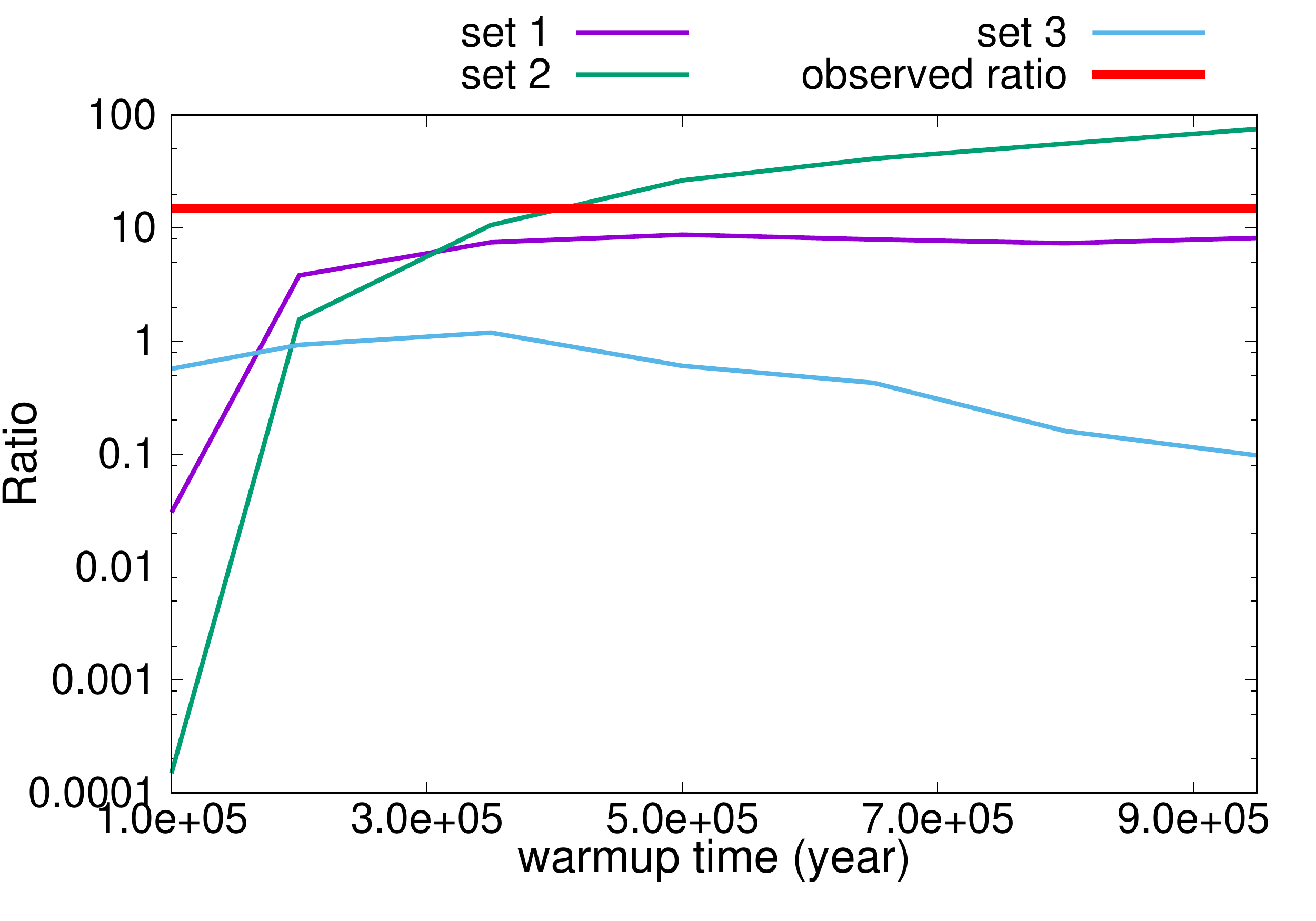}
    \includegraphics[width=5 cm, angle= 270]{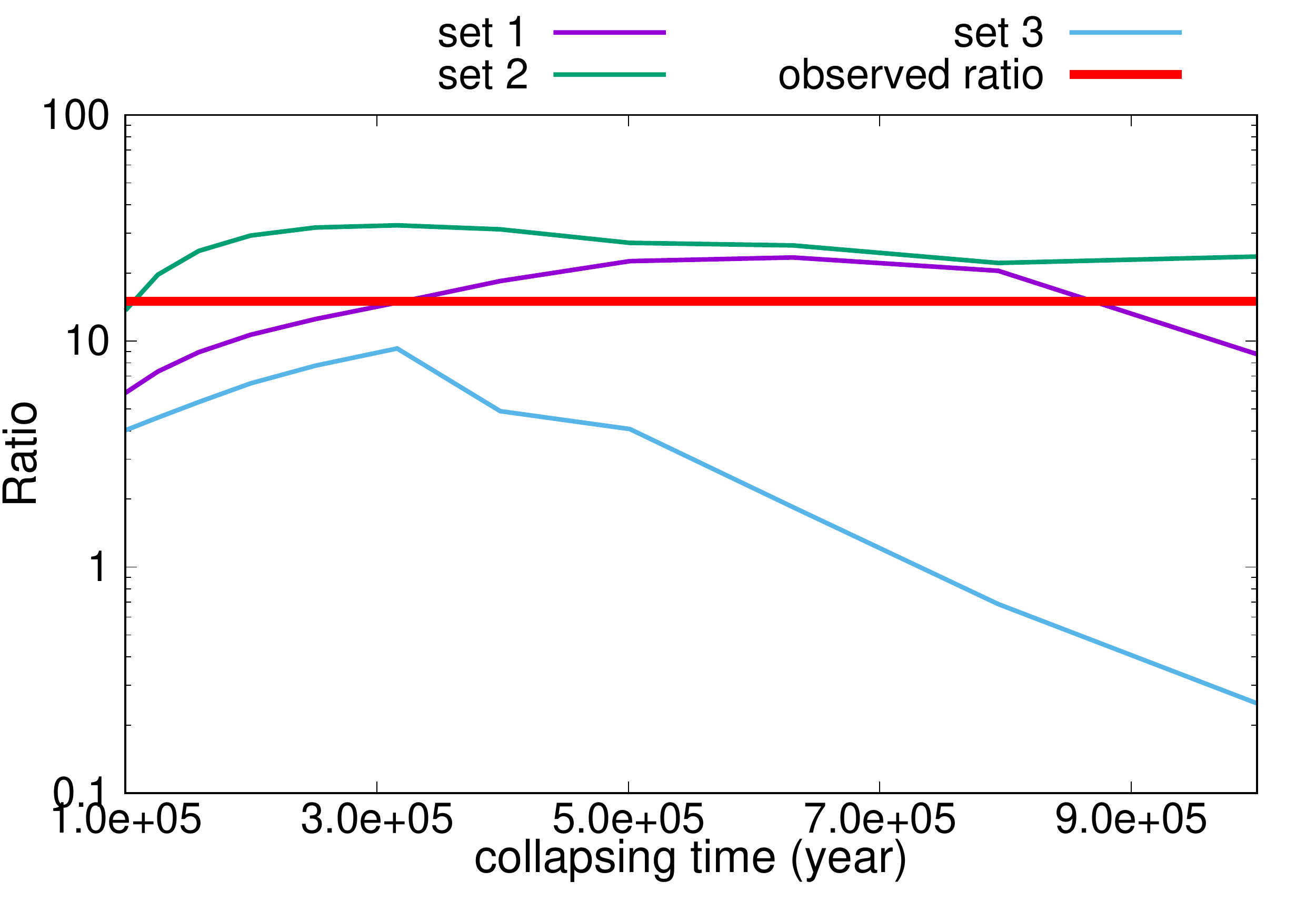}
    \caption{Peak abundance ratio between $\rm{C_2H_5CN}$ and $\rm{CH_2CHCN}$ for various BE sets. The left panel shows results for Model A and right panel for Model B. The observed ratio is shown with the horizontal solid red line.}
    \label{fig:rat_EV}
\end{figure*}

An exciting trend for $\rm{CH_2CHCN}$ is obtained when we vary the warm-up time of Model A. We notice that the peak abundance value of $\rm{CH_2CHCN}$ is gradually shifted towards the higher temperature with the decrease in the warm-up time scale. For example, for set 1, with $\rm{t_{warm1}}=10^6$ years, we get a peak abundance at 146 K (117 K for set 2 and 145 K for set 3), which is shifted to 197 K (143 K for set 2 and 161 K for set 3) with $\rm{t_{warm1}}=10^5$ years. 
Fig. \ref{fig:EV} shows that the ice-phase abundance of $\rm{CH_2CHCN}$ declines around 70 K. However, its peak abundance appears at a much higher temperature. It is because of the involvement of the gas phase pathways for the formation of $\rm{CH_2CHCN}$. In the $\rm{C_2H_5CN}$, the peak abundance obtained remains roughly invariant with the variation of the warm-up time scale (varies in the range $108-117$ K for set 1, $99-101$ K for set 2, and $95-103$ for set 3). \cite{garr17} obtained $129-131$ K for the different warm-up time scales.

\cite{bell16} obtained an abundance ratio between $\rm{C_2H_5CN}$ and $\rm{CH_2CHCN}$ of $\sim 15$ in Sgr B2(N2). The obtained ratio from our various models and various sets of BEs are noted in Table \ref{tab:ratios}.
The peak abundance of $\rm{C_2H_5CN}$ and $\rm{CH_2CHCN}$ does not appear simultaneously, but we consider their peak values in deriving this ratio. So, there should be some uncertainty in the derived molecular ratio.
For set 1 BE, the values noted in Table \ref{tab:pk-abn}, show a ratio between the peak abundance of $\rm{C_2H_5CN}$ to $\rm{CH_2CHCN}$ as $7.2$. For more higher warm-up time scale ($\rm{t_{warm1}}=5 \times 10^5$ years), it can goes up to $\sim$ 9 (see Fig. \ref{fig:rat_EV}). 
The right panel of Fig. \ref{fig:rat_EV} shows the ratio obtained by varying the collapsing time (Model B). 
Overall, with the variation of warm-up ($t_{warm1}=2 \times 10^5 - 10^6$ years) and collapsing time ($t_{coll}=10^5-10^6$ years) shown in Fig. \ref{fig:rat_EV}, we obtain a ratio of $3.8-23.4$ for set 1. 
For the similar time scales, this ratio varies in the range of $1.56-72$ and $0.8-9.3$ for set 2 and set 3, respectively.
We notice a abrupt decrease in ratio for set 1 and set 2 for a shorter warm-up time scale ($t_{warm1}=10^5$ years). It is 0.033 and 0.00014 for set 1 and 2, respectively.
A vast difference between the ratio of sets 2 and 3 is observed in both panels of Fig. \ref{fig:rat_EV}. It happens mainly because of the changes in the adsorption energies of the H atom. For set 3, due to the higher adsorption energy of the H atom, it has a longer residence time on the grain, which eventually might helps in the formation of more saturated species.

\begin{figure*}
    \centering
    \includegraphics[width=10cm, angle= 270]{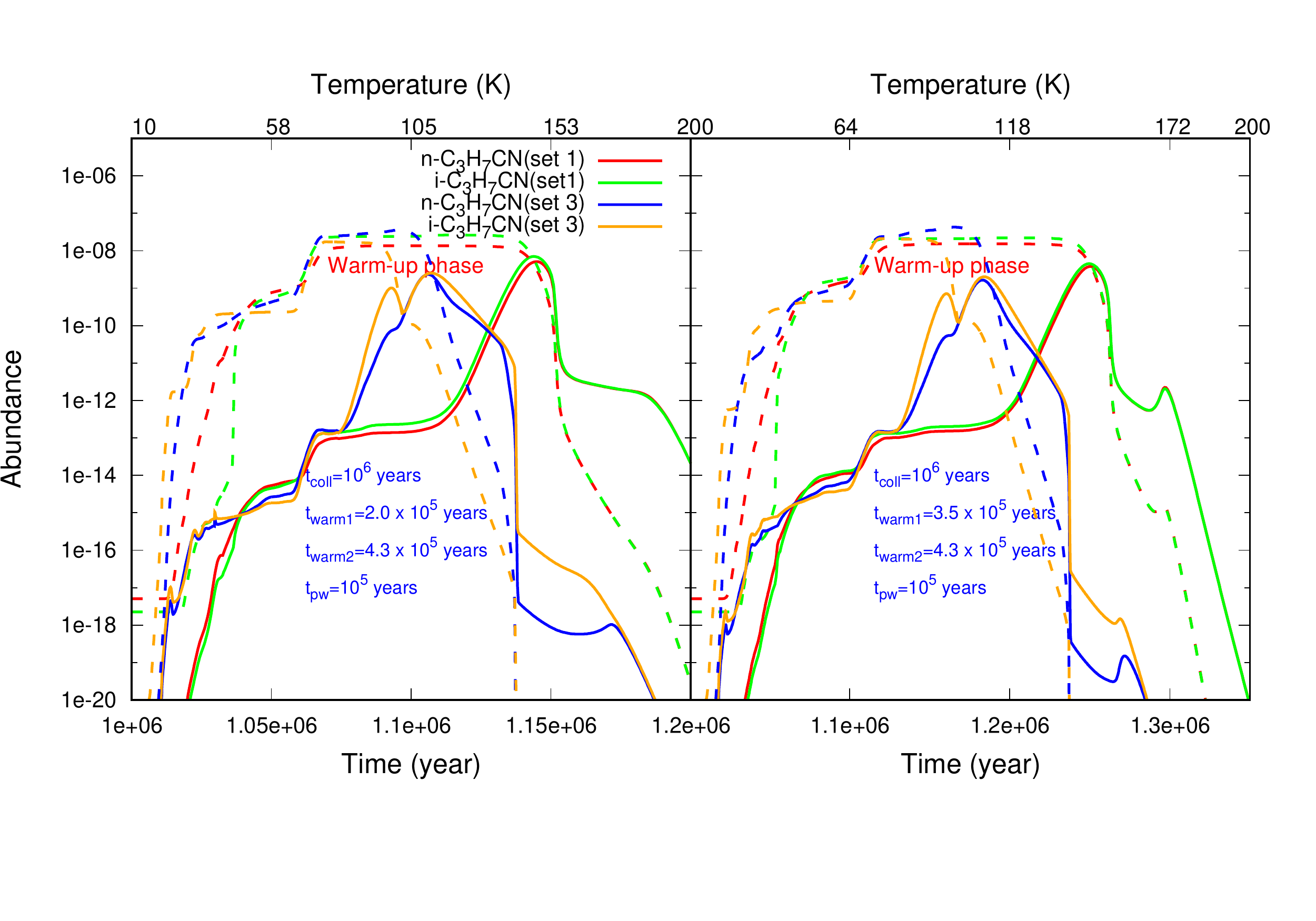}
    \caption{Time evolution of $\rm{n-C_3H_7CN}$ and ${\rm i-C_3H_7CN}$ in the warm-up phase for set 1 and set 3 BEs. Solid curves represent the gas-phase abundance, whereas the dashed curves represent the abundance of ice-phase species.}
    \label{fig:c3h7cn}
\end{figure*}

\begin{figure*}
    \centering
    \includegraphics[width=5cm, angle= 270]{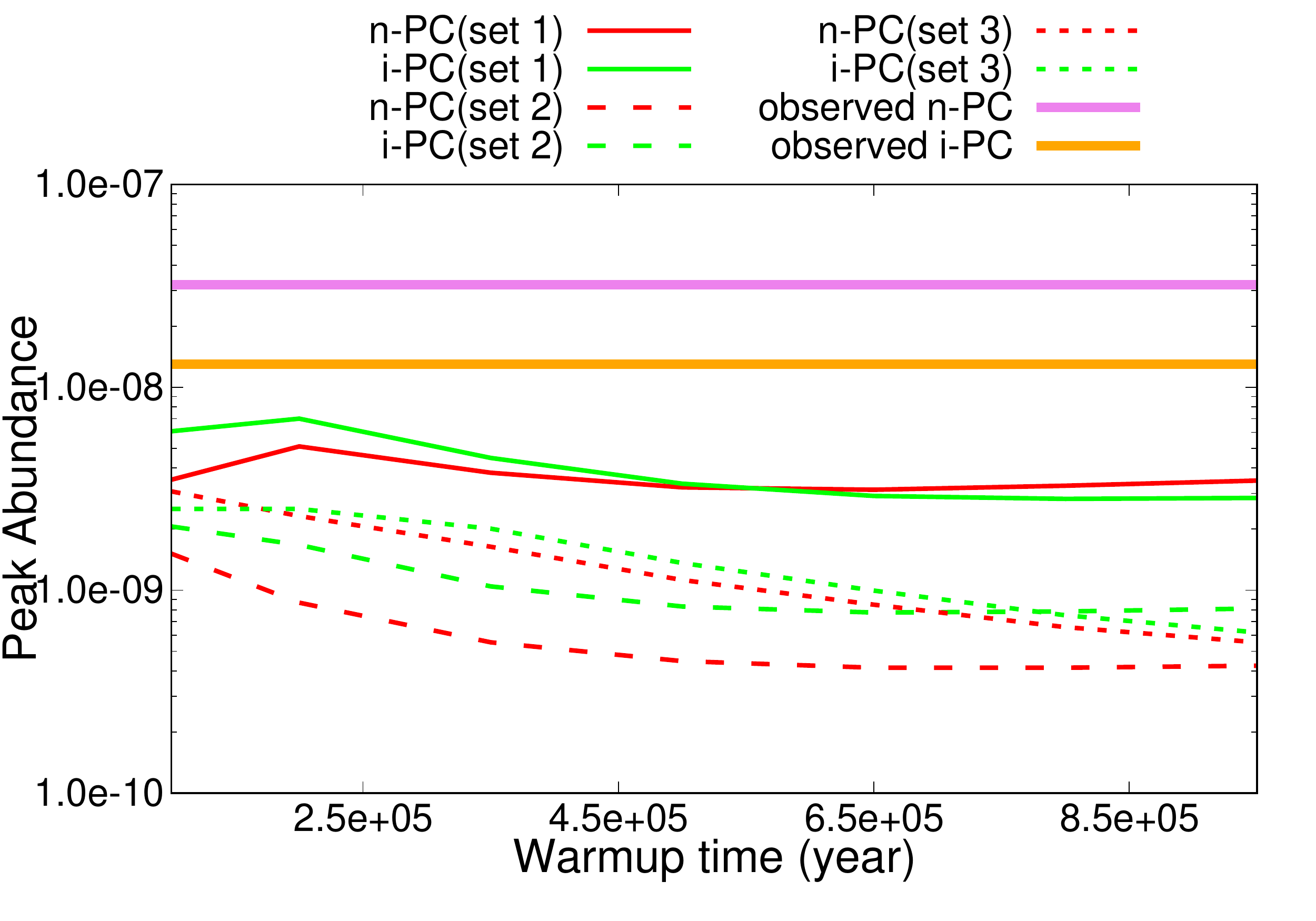}
    \includegraphics[width=5cm, angle= 270]{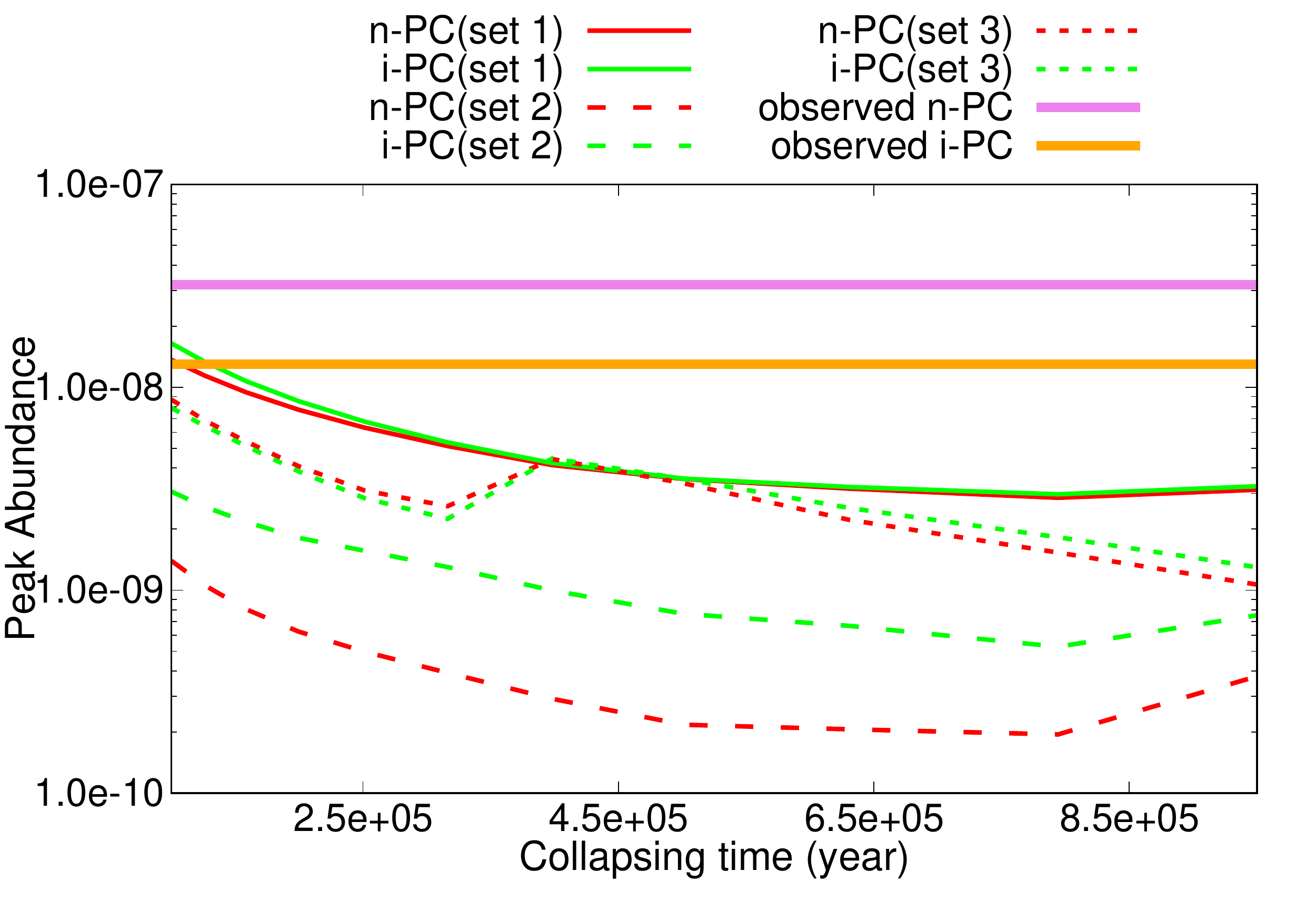}
    \caption{Peak abundance of $\rm{n-C_3H_7CN}$ (n-PC) and $\rm{i-C_3H_7CN}$ (i-PC) for various BE sets. The left panel shows results for Model A and right panel for Model B. Observed peak abundances of n-PC and i-PC in Sgr B2(N) are also shown with the horizontal solid lines.}
    \label{fig:PC-warm-coll}
\end{figure*}

 \begin{figure*}
    \centering
    \includegraphics[width=5cm, angle= 270]{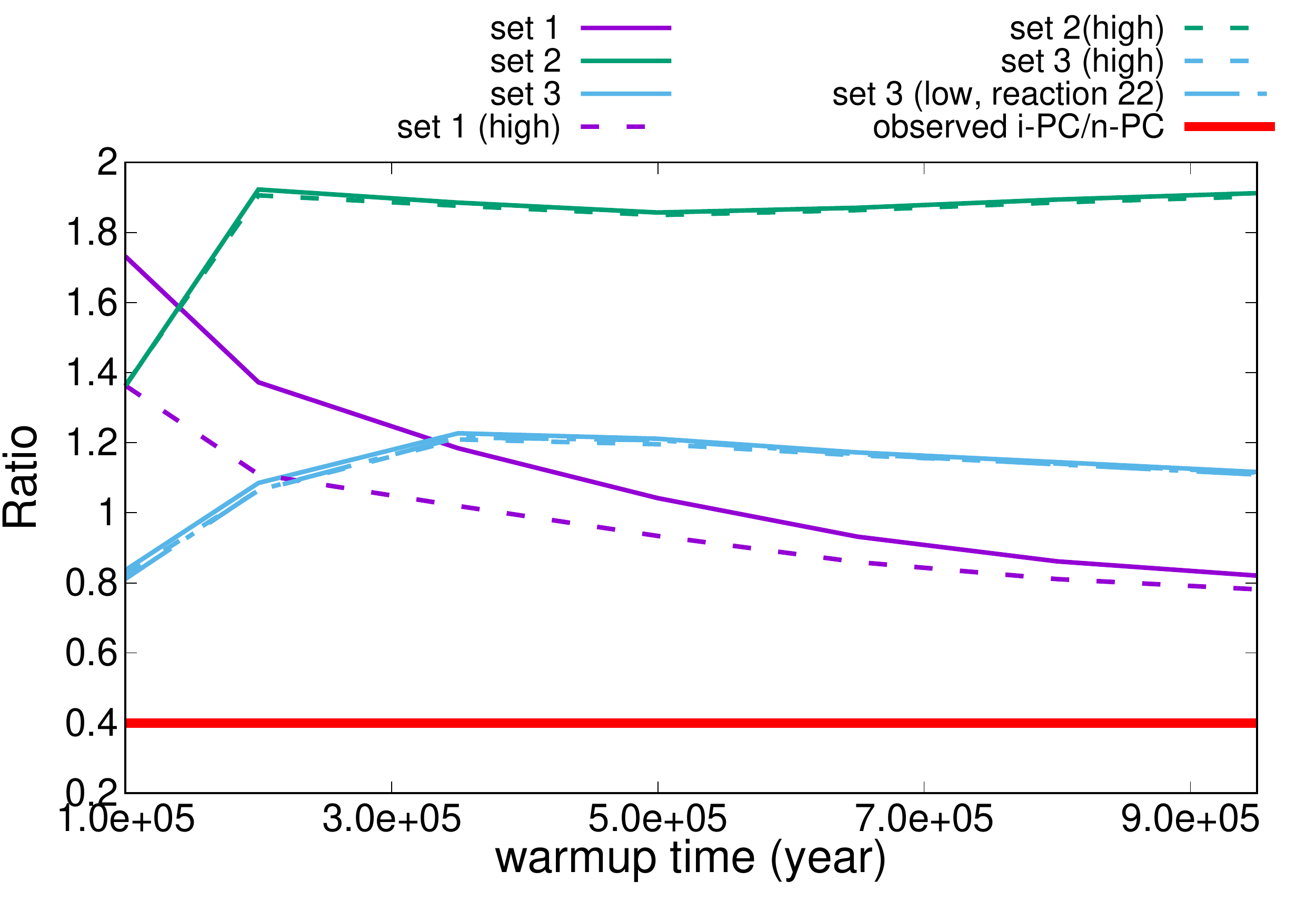}
     \includegraphics[width=5cm, angle= 270]{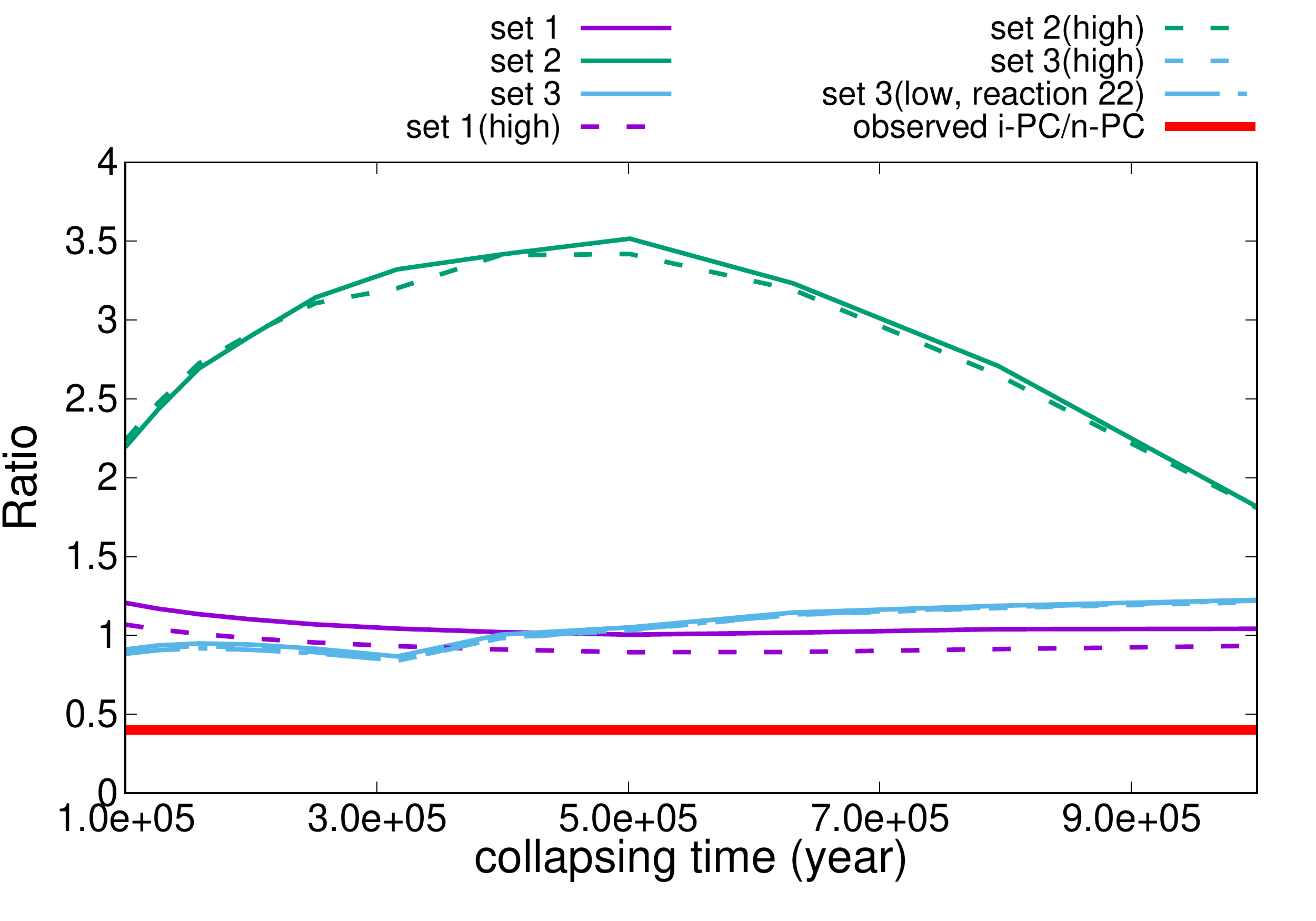}
    \caption{Peak abundance ratio of $\rm{C_3H_7CN}$ (i/n) for various BE sets. The left panel shows results for Model A and right panel for Model B. The ratios considering the high activation barriers are shown with the dashed curve. For set 1, the i/n ratio deviate largely from the low barriers for the variation of warm-up time scale. The observed value is shown with the solid red horizontal line.}
    \label{fig:PC_rat}
\end{figure*}

\begin{figure*}
    \centering
    \includegraphics[width=5cm, angle= 270]{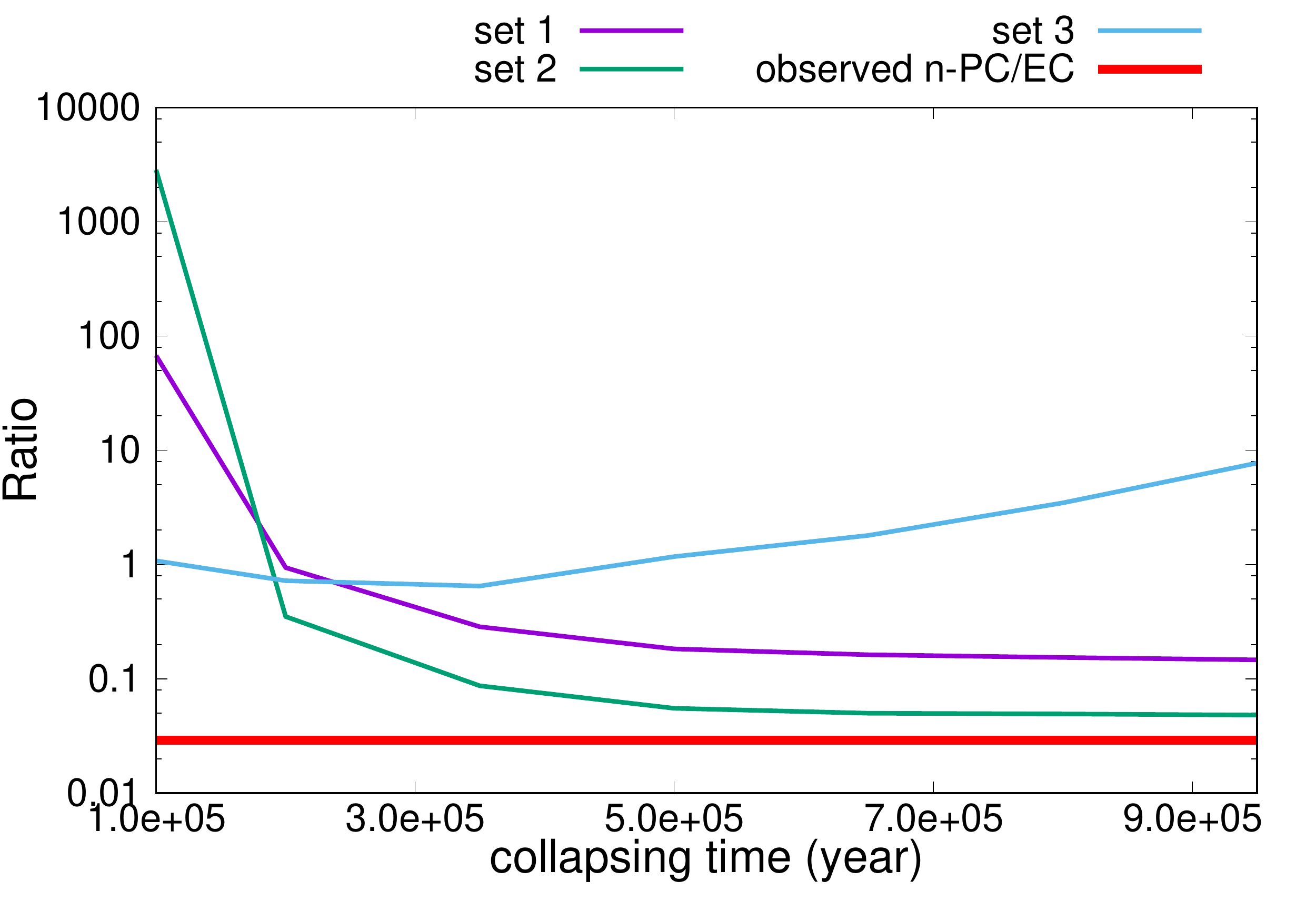}
     \includegraphics[width=5cm, angle= 270]{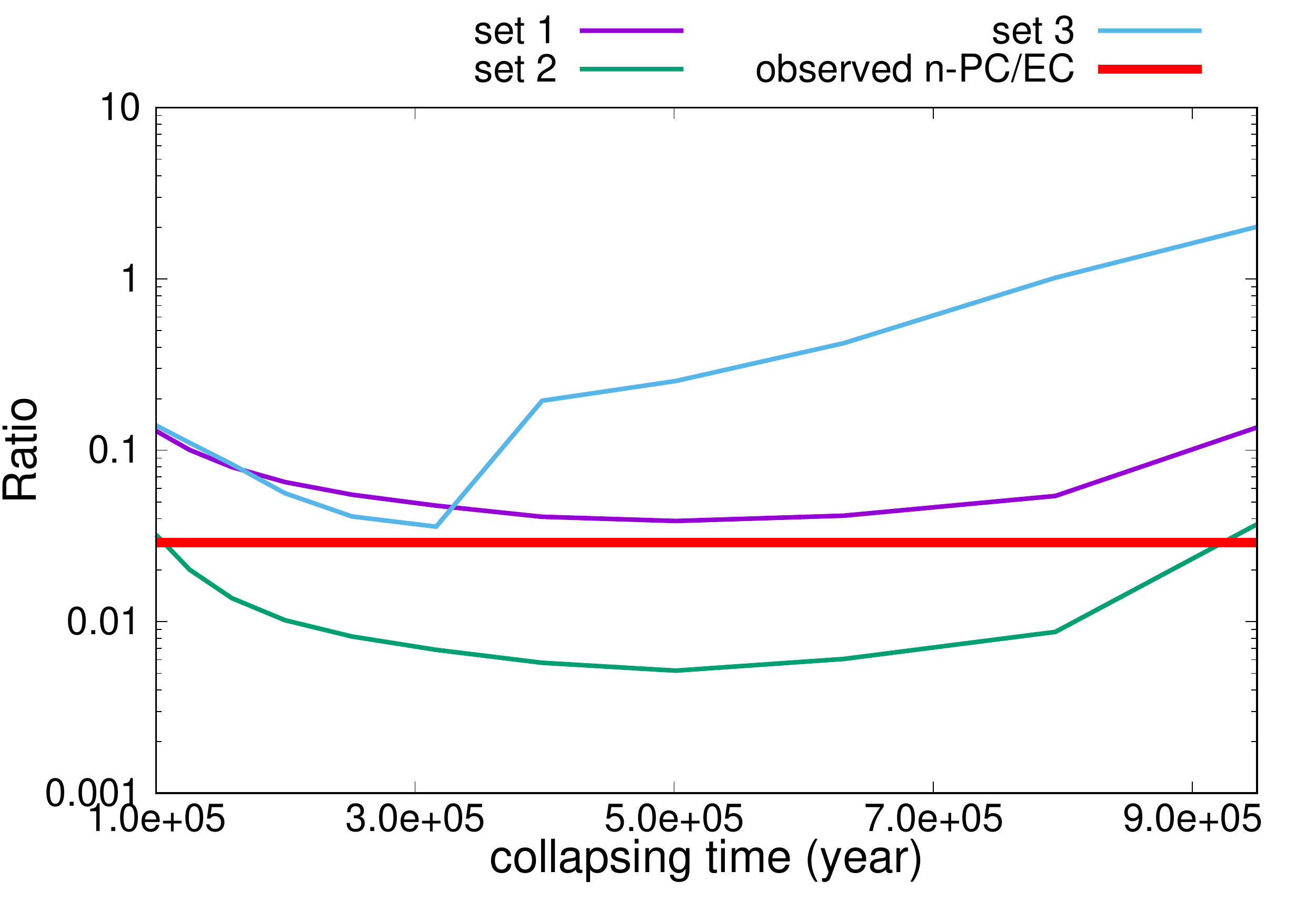}
    \caption{Peak abundance ratio between $\rm{n-C_3H_7CN}$ and $\rm{C_2H_5CN}$. The left panel shows results for Model A and right panel for Model B. The observed value is shown with the red horizontal line.}
    \label{fig:PC_Eth-rat}
\end{figure*}

\begin{figure*}
    \centering
    \includegraphics[width=5cm, angle= 270]{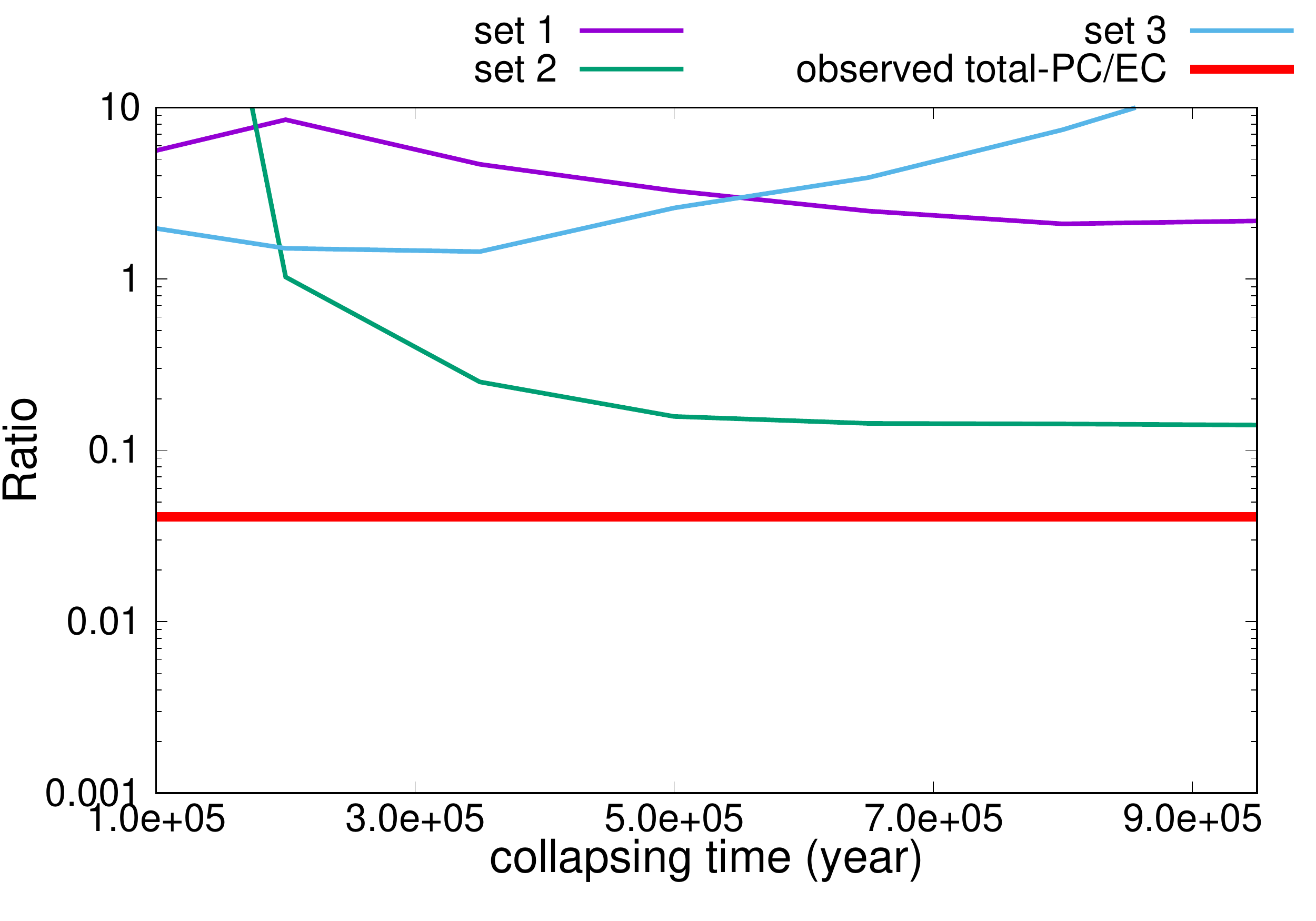}
     \includegraphics[width=5cm, angle= 270]{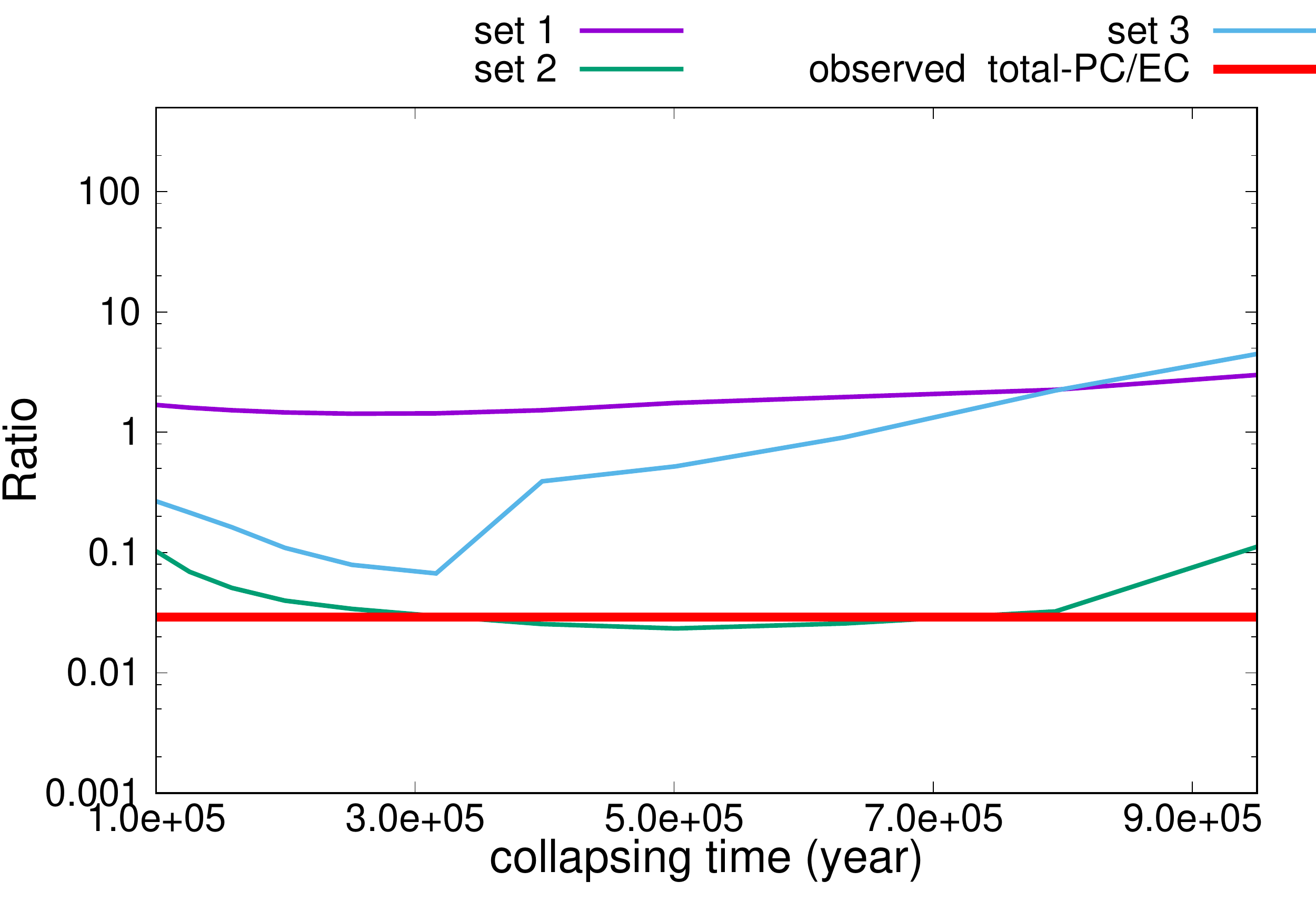}
    \caption{Peak abundance ratio between total $\rm{C_3H_7CN}$ and $\rm{C_2H_5CN}$ for various BE sets. The left panel shows results for Model A and right panel for Model B. The observed value is shown with the red horizontal line.}
    \label{fig:tot-PC_Eth-rat}
\end{figure*}

\begin{figure*}
    \centering
\includegraphics[width=10cm, angle= 270]{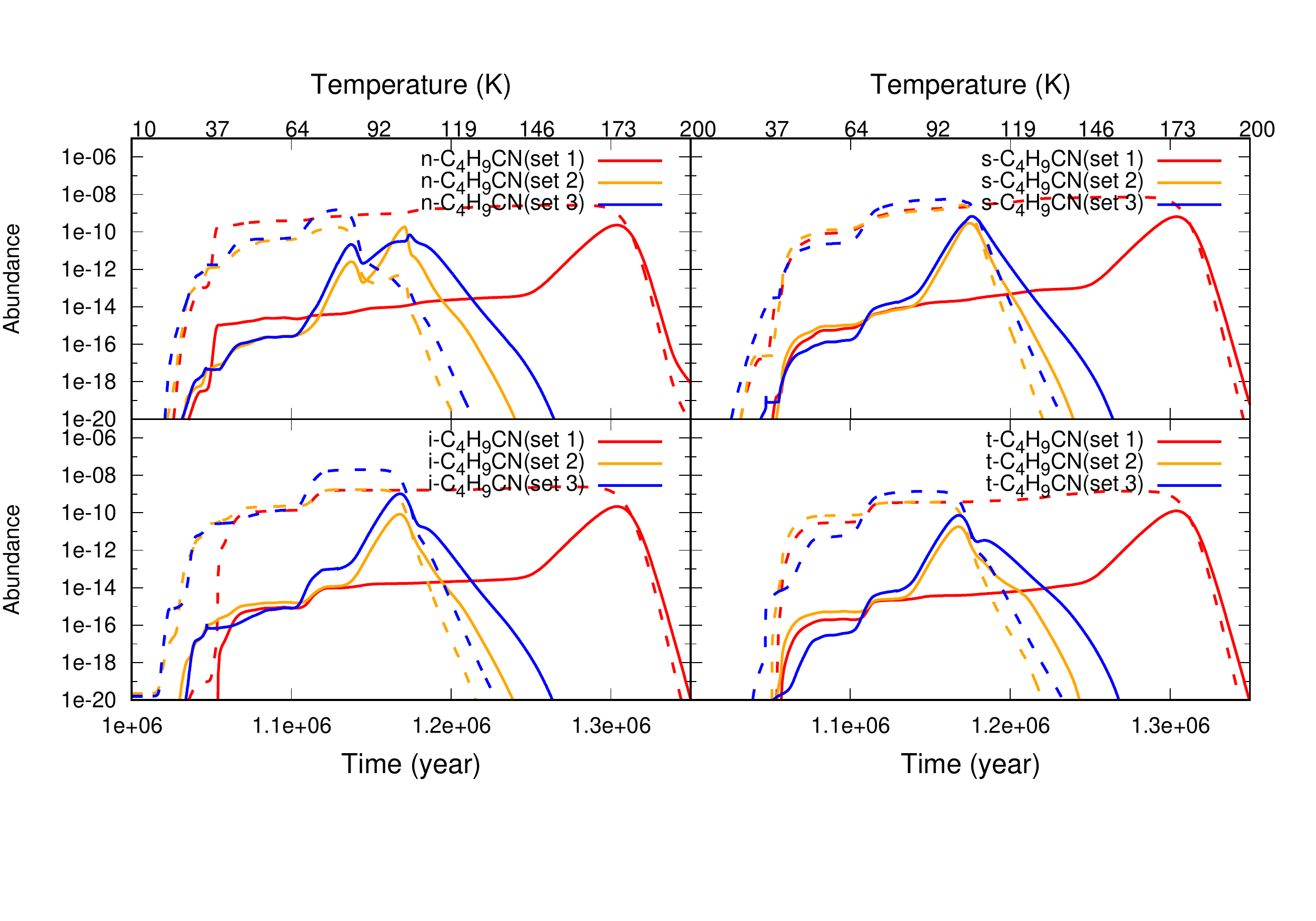}
    \caption{Time evolution of $\rm{n-C_4H_9CN}$, $\rm{i-C_4H_9CN}$, $\rm{s-C_4H_9CN}$, and $\rm{t-C_4H_9CN}$ during the warm-up phase for various BE sets. Solid curves represent the gas-phase abundance, whereas the dashed curves represent the abundance of ice-phase species.}
    \label{fig:c4h9cn}
\end{figure*}

\begin{figure*}
    \centering
    \includegraphics[width=8cm, angle= 270]{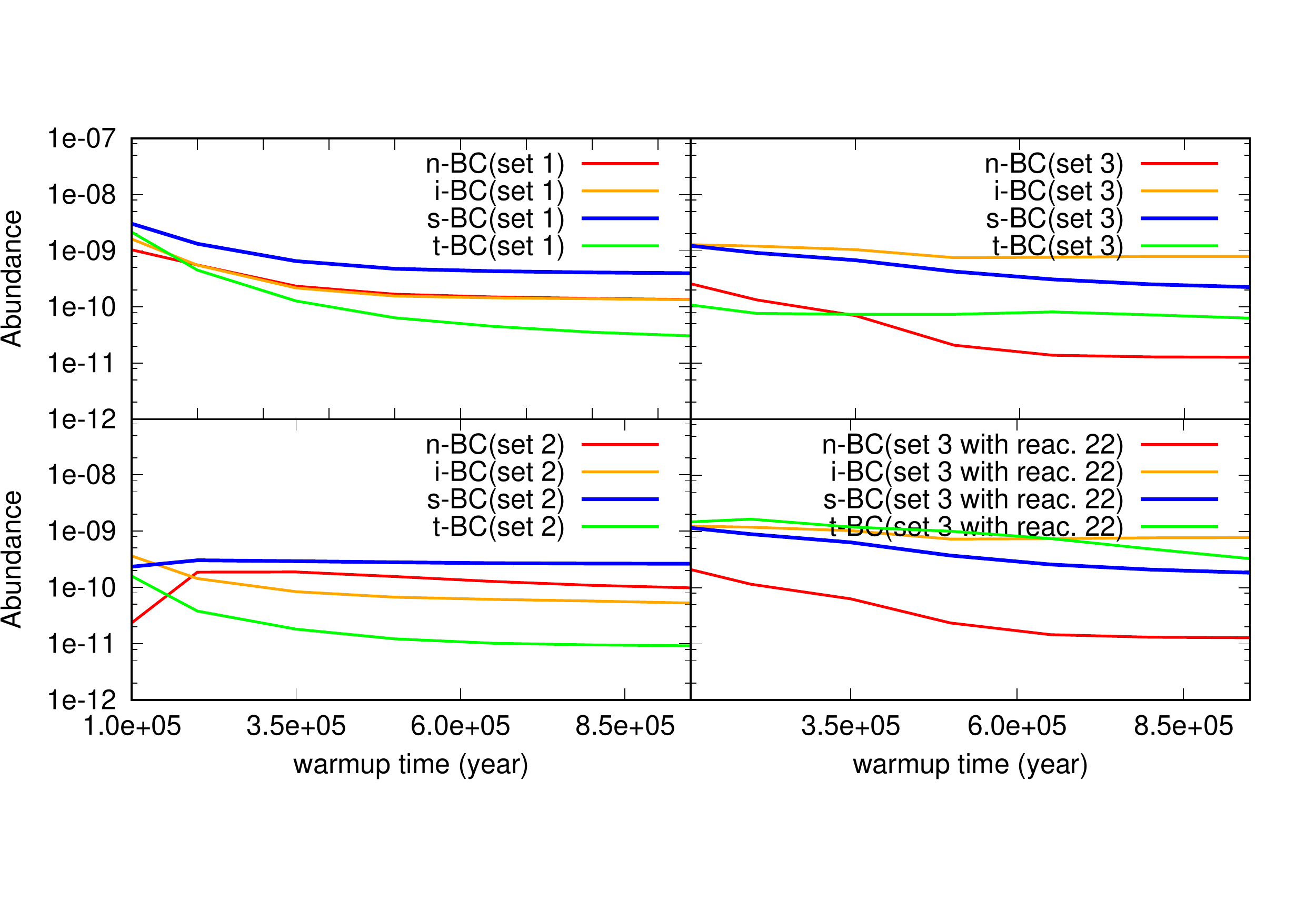}
    \includegraphics[width=8cm, angle= 270]{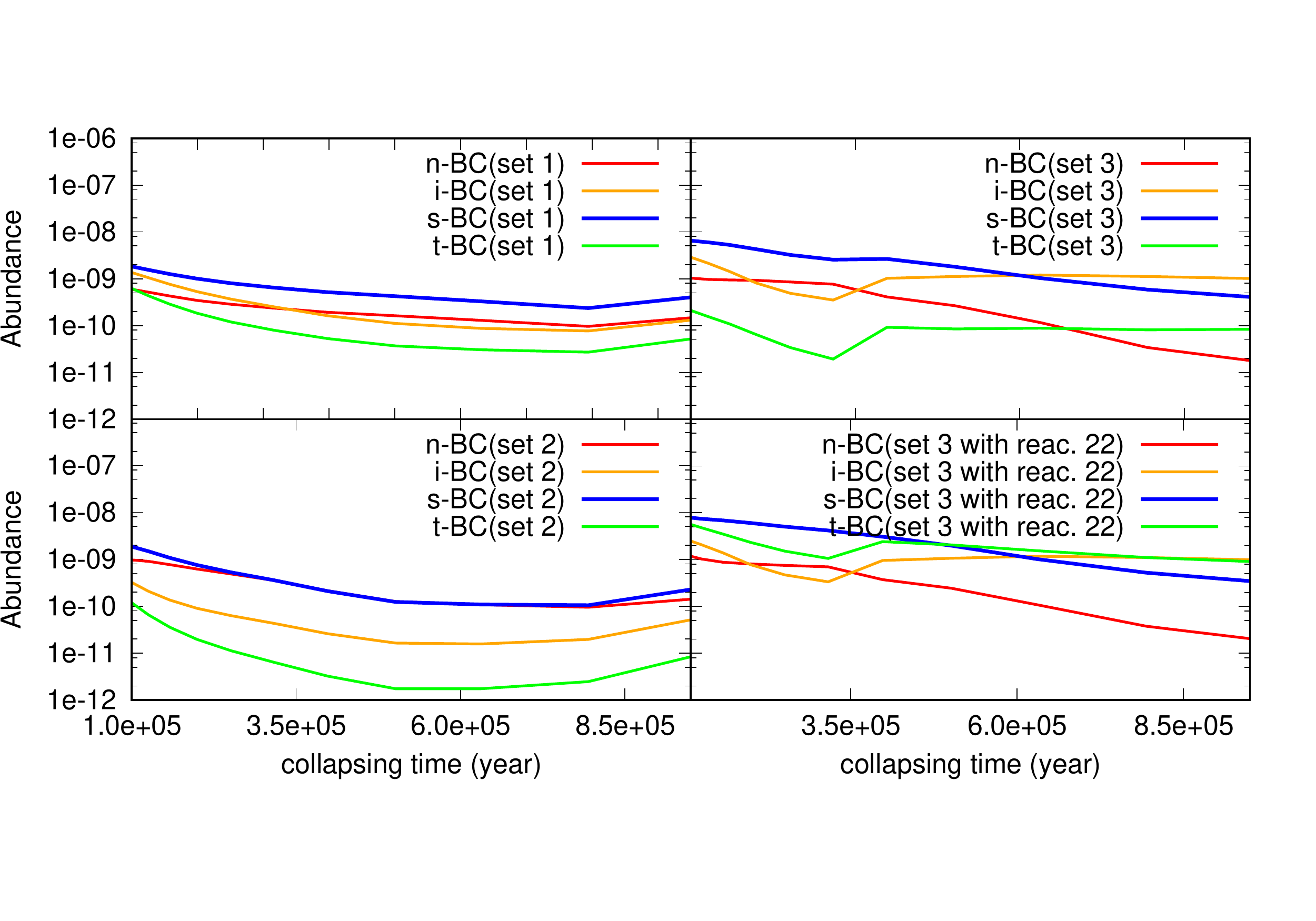}
    \caption{Peak abundances of $\rm{C_4H_9CN}$ (n,i,s, and t) for various BE sets. The upper panel shows results for Model A and lower panel for Model B.}
    \label{fig:collapse-butyl}
\end{figure*}

\begin{figure*}
    \centering
    \includegraphics[width=5.0cm, angle= 270]{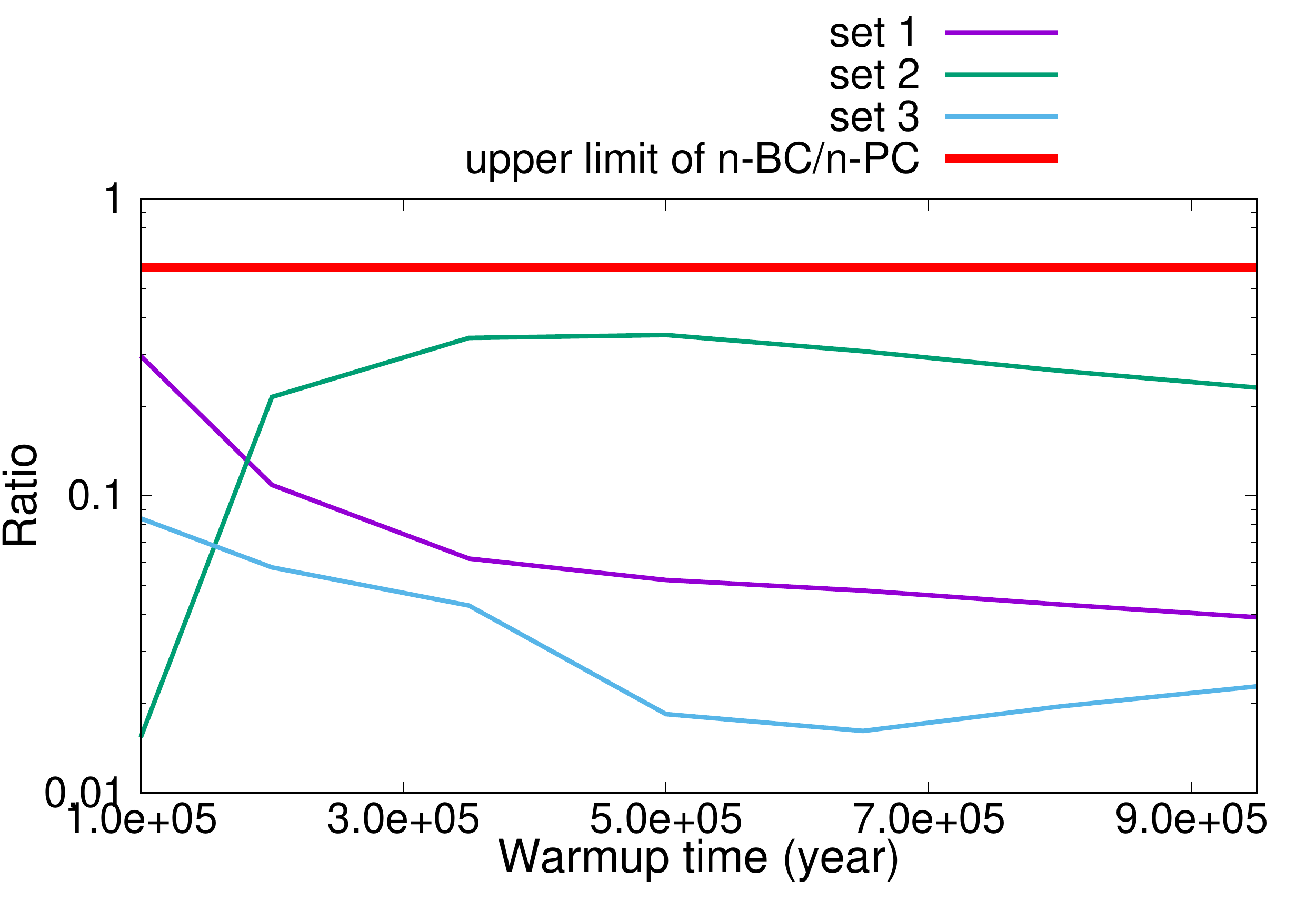}
     \includegraphics[width=5.0cm, angle= 270]{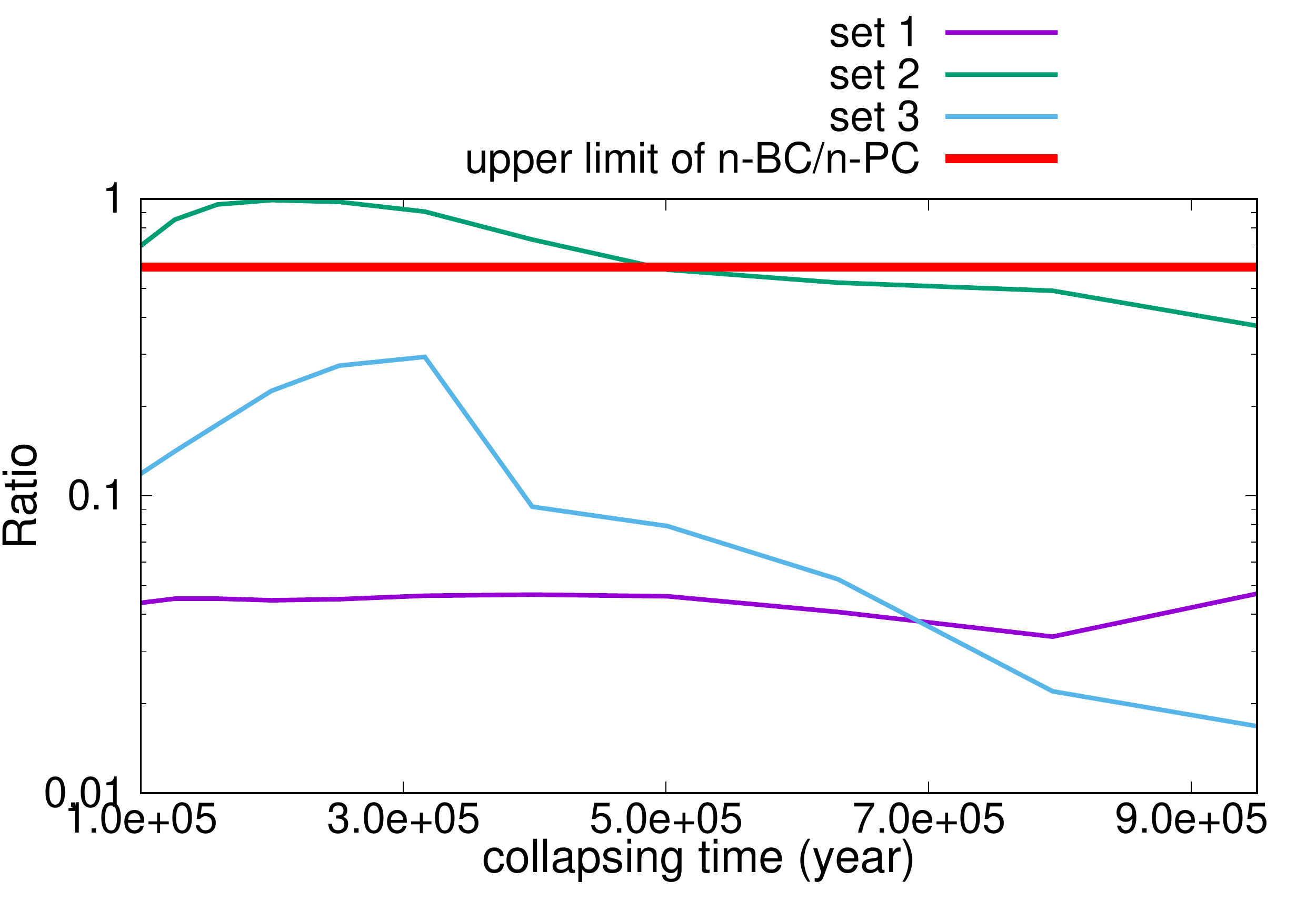}
    \caption{Peak abundance ratio between $\rm{n-C_4H_9CN}$ and $\rm{n-C_3H_7CN}$ for various BE sets. The left panel shows results for Model A and right panel for Model B. Derived upper-limit of this ratio \citep[0.59,][]{garr17} is shown with the red horizontal line.}
    \label{fig:but_pc-rat}
\end{figure*}

\begin{figure*}
   \centering
    \includegraphics[width=5cm, angle= 270]{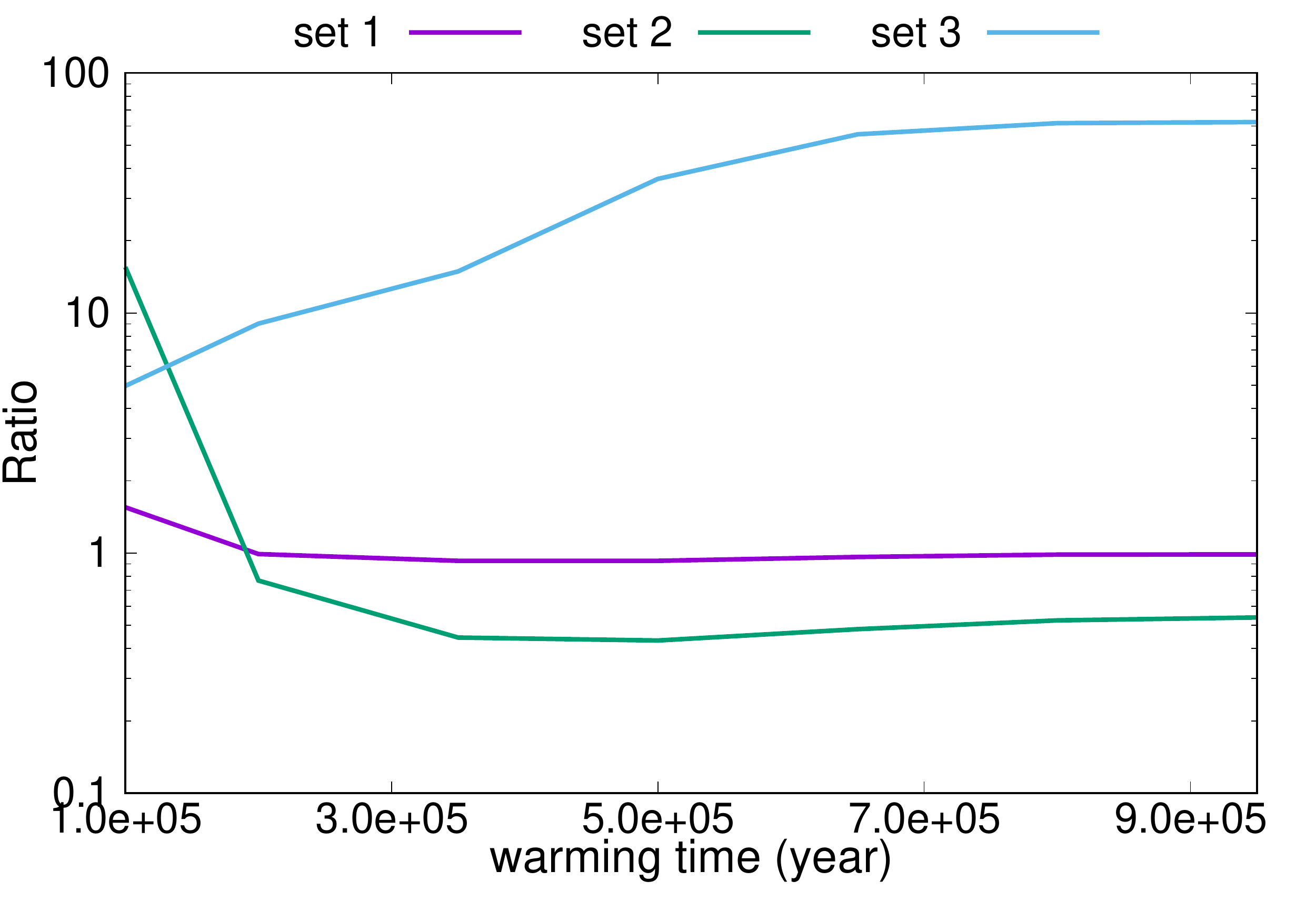}
    \includegraphics[width=5cm, angle= 270]{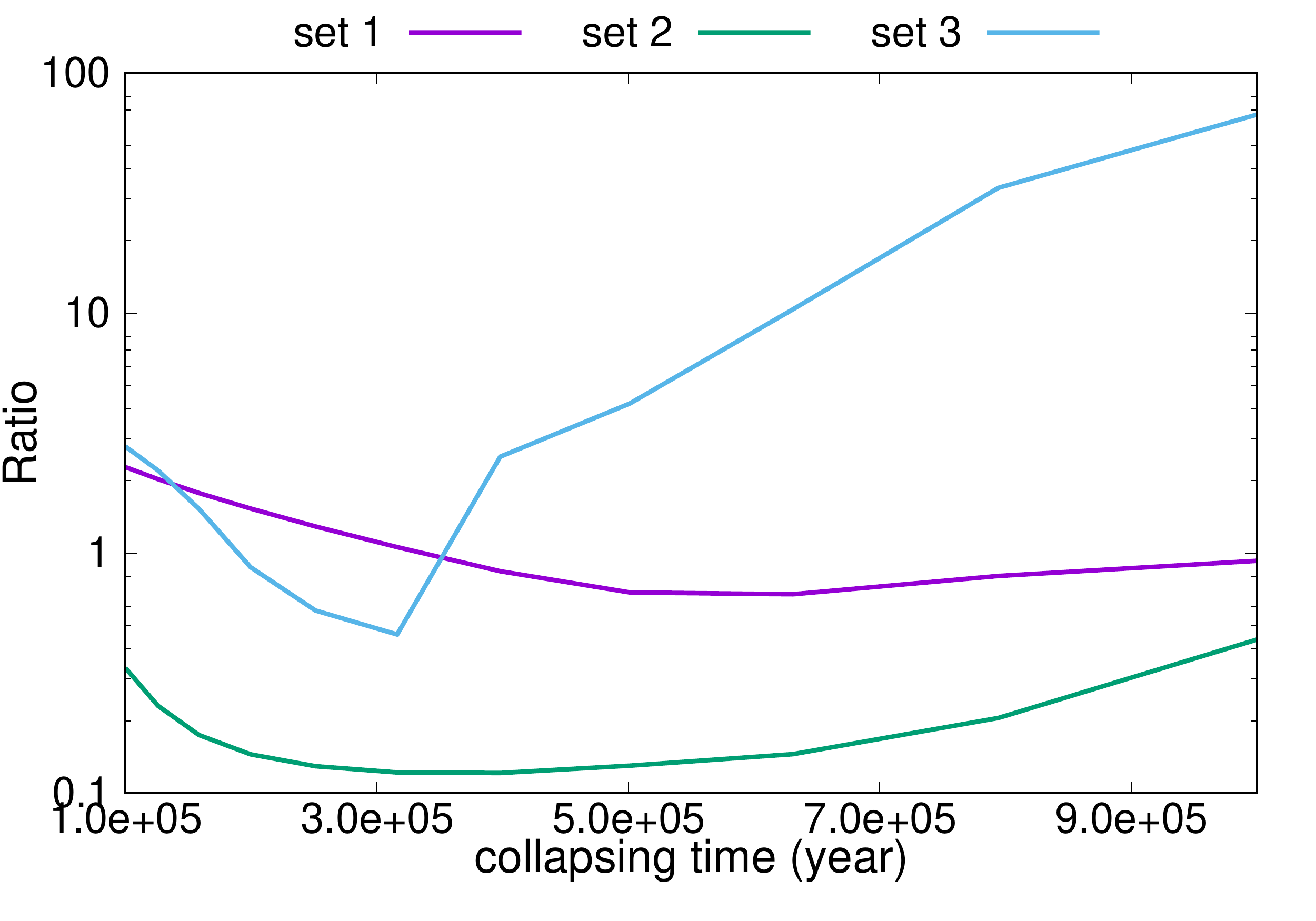}
    \caption{Peak abundance ratio between $\rm{i-C_4H_9CN}$ and $\rm{n-C_4H_9CN}$ for various BE sets. The left panel shows results for Model A and right panel for Model B.}
    \label{fig:in-butyl}
\end{figure*}

\subsection{Propyl cyanides}
Formation pathways of $\rm{i/n-C_3H_7CN}$ are discussed in section \ref{sec:prop}. \cite{garr17} considered the BE of both the isomers the same (7237 K). We obtain relatively lower BE values of these two species with the monomer and tetramer configuration. With the monomer substrate, a  higher BE for $\rm{i-C_3H_7CN}$ (3279 K) than $\rm{n-C_3H_7CN}$ (2991 K) is obtained. On the contrary, with the tetramer configuration, we obtain an opposite trend (4970 K for $\rm{i-C_3H_7CN}$ and 5567 K for $\rm{n-C_3H_7CN}$), which are much lower than the values used in \cite{garr17}.

Fig. \ref{fig:c3h7cn} shows the time evolution of the two forms of $\rm{C_3H_7CN}$ (i and n) by considering $\rm{t_{warm1}=2 \times 10^5}$ years and $\rm{t_{warm1}=3.5 \times 10^5}$ years, respectively with Model A (with set 1 and set 2 BEs). With the warm-up time scales, we notice a significant difference in the abundance of $\rm{C_3H_7CN}$. In general, a shorter warm-up time yields a larger peak abundance (see the left panel of Fig. \ref{fig:PC-warm-coll}). 
For $\rm{t_{warm1}}>6 \times 10^5$ years, we obtain i$<$n, whereas the opposite is valid for a shorter warm-up time scale for set 1. For the set 2 BE, we always see i$>$n. For set 3 also, we obtain i$>$n except for the shortest warm-up time ($10^5$ years). The right panel of Fig. \ref{fig:PC-warm-coll} shows the peak abundances of n and i $\rm{C_3H_7CN}$ for the variation in the collapsing time for set 1, set 2, set 3. It shows that for set 1, peak gas-phase abundance of $\rm{i-C_3H_7CN}$ is always greater ($2.97 \times 10^{-9}-1.65 \times 10^{-8}$) than $\rm{n-C_3H_7CN}$ ($2.86 \times 10^{-9 }-1.36 \times 10^{-8}$). For set 2, we always have obtained a high abundance of $\rm{i-C_3H_7CN}$ than $\rm{n-C_3H_7CN}$. For set 3, i$>$n is obtained when ${\rm t_{coll}> 5 \times 10^5}$ years and i$<$n for the shorter ${\rm{t_{coll}}}$.
 From various models (high/low) barriers and with various warm-up time scales), \cite{garr17} obtained the abundance $1.9 \times 10^{-10}-46.0 \times 10^{-10}$, and $7.6 \times 10^{-10}-34 \times 10^{-10}$ for n and i $\rm{C_3H_7CN}$, respectively. \cite{bell14} calculated the abundance of $3.6 \times 10^{-9}$ and $8.0 \times 10^{-9}$ for n and i conformers of $\rm{C_3H_7CN}$, respectively, whereas from the observations, they obtained abundance of $3.2 \times 10^{-8}$ and $1.3 \times 10^{-8}$, respectively. These observed values are shown in Fig. \ref{fig:PC-warm-coll} with horizontal lines for better understanding. However, as like \cite{bell14,garr17}, none of our models were successful in obtaining such a high abundance with the present reaction network. It might be due to the exceptional environment in the Galactic center region than the other not core as discussed in \cite{bonf19,will20}.

The i/n abundance ratio for set 1 with low activation barriers vary between 1 and 1.2 (see Fig. \ref{fig:PC_rat}). With the slow and low barrier model, \cite{garr17} obtained a peak abundance of $1.9 \times 10^{-9}$ and $3.4 \times 10^{-9}$, respectively for $\rm{n-C_3H_7CN}$ and $\rm{i-C_3H_7CN}$, which yields an i/n abundance ratio of 1.79. Fig. \ref{fig:PC_rat} represents the obtained peak ratio between i and n $\rm{C_3H_7CN}$ with the warm-up time scale (left) and collapsing time scale (right). The ratio obtained with other BEs are noted in Table \ref{tab:ratios}.
\cite{bell14,bell16} observed the i/n ratio of $\sim 0.4 \pm0.06$. None of our models with the low activation barriers (for the reaction of CN with C$_2$H$_2$, C$_2$H$_4$, C$_3$H$_6$, and C$_4$H$_8$) were able to achieve the observed value (either Model A or Model B). With the high activation barrier, we found that this ratio varies between 0.8 and 1.4 for set 1. 
In the case of set 2 and set 3, mostly we have i$>$n. With set 3, we obtain $i/n \sim 0.7$ when a shorter warm-up  ($t_{warm1}=10^5$ years for Model A) and shorter collapsing time ($t_{coll}=1-3 \times 10^5$ years for Model B) are used. One point to be remembered that the ratio itself is time-dependent, and we consider the ratio with their respective peak values. 
\cite{bell16} also observed $\rm{n-C_3H_7CN/C_2H_5CN}=0.029$ and $\rm{total \ C_3H_7CN/C_2H_5CN}=0.041$. 
 Figs. \ref{fig:PC_Eth-rat} and \ref{fig:tot-PC_Eth-rat} represent the simulated peak ratio of $\rm{n-C_3H_7CN/C_2H_5CN}$ and $\rm{total \ C_3H_7CN/C_2H_5CN}$, respectively. These ratios are falling in between our simulated range.
 
In section \ref{sec:prop}, reactions related to the formation of $\rm{C_3H_7CN}$ are discussed. The TS calculation of reaction 22 reveals an activation barrier of $947$ K. So far, we do not consider the destruction of $\rm{i-C_3H_7CN}$ by any hydrogenation reaction in our network. By considering reaction 22, we observe a few changes in the abundances of $\rm{i-C_3H_7CN}$ because of the formation by reaction 19 again. However, its effect on $\rm{C_4H_9CN}$ is vital and discussed in the latter portion. The dot-dashed curve represents the peak abundance ratio between i and n $\rm{C_3H_7CN}$ (low barrier) in Fig. \ref{fig:PC_rat}. 
No significant changes in the abundance of $\rm{i-C_3H_7CN}$ are seen with the inclusion of reaction 22. It is because a substantial part of the product of reaction 22 ($\rm{CH_3(C)CH_3CN}$) hydrogenate to form $\rm{i-C_3H_7CN}$ by reaction 19 again.

\subsection{Butyl cyanide}
For the formation of various forms of $\rm{C_4H_9CN}$, we consider ice phase reactions $23-36$ of section \ref{sec:BC}. Fig. \ref{fig:c4h9cn} shows the time evolution of n/i/s/t-C$_4$H$_9$CN for various BEs. \cite{garr17} used the BE of these four species as 8937 K. We obtain significantly lower BE values with our calculations. With the monomer configuration, we obtain a higher BE (scaled by 1.416) for n (6320 K) followed by t (4574 K), i (4422 K), and s (2635 K). However, with the larger substrate (tetramer), we obtain the highest BE (scaled by 1.188) for s (5313 K), followed by i (5151 K), t (5148 K), and n (4388 K). With the set 1 BE, we obtain the peak values of these species at 175 K. With the tetramer structure (set 2 and set 3), due to lower BEs, the peak value of these species appears to be around $100-105$ K.

Fig. \ref{fig:collapse-butyl} shows the effect on the peak abundances of various forms of $\rm{C_4H_9CN}$ for the variation of warm-up time (top) and collapsing time (bottom). In general, it seems that the shorter collapsing time and shorter warm-up time increase the production of all these species. \cite{garr17} used the set 1 BE values with $\rm{t_{coll}}=10^6$ years. With the shortest warm-up time ($\sim 5 \times 10^4$ years with low and high barriers), they obtained the highest peak abundance of s, followed by n, i, and t, and with the largest warm-up time ($\sim 10^6$ years), they obtained the highest peak abundance of i, followed by s, n, and t. So, in all the cases, they got the lowest abundance of t. In our set 1 model (with a low barrier and ${\rm t_{coll}=10^6}$ years) for most of the warm-up time scale, the sequence is s, n, i, and t.

We obtain a dramatic change in the abundances with set 2 and set 3 BE values. For set 3, where the most updated BE values are used, we obtain a peak abundance sequence (with $\rm{t_{coll}=10^6}$ years) of i, s, t, and n for most of the warm-up time (see at the top of Fig. \ref{fig:collapse-butyl}). For the shorter warm-up time ($\sim 10^5$ years), t and n interchange their position. This model has considered $\rm{t_{coll}}=10^6$ years. For the shorter collapsing time, s has the maximum abundance, whereas t has the minimum abundance. 

With the consideration of the reaction 22, scenario has changed dramatically. With ${\rm t_{coll}=10^6}$ years, when we consider reaction 22, for low barrier case, we have an abrupt increase in the abundance of the $\rm{t-C_4H_9CN}$ (last panel of the top of Fig. \ref{fig:collapse-butyl}). For example, Table \ref{tab:pk-abn} shows the abundance of $\rm{t-C_4H_9CN}$ (with $\rm{t_{coll}}=10^6$ years and $\rm{t_{warm1}=3.5 \times 10^5}$ years) is $7.37 \times 10^{-11}$ when reaction 22 is not considered. On the contrary, with reaction 22, we have its abundance $1.29 \times 10^{-9}$. So, an increase of two orders of magnitude is obtained.
It happens due to the production of $\rm{CH_3(C)CH_3CN}$ by reaction 22, which is further used in reaction 36 for the formation of $\rm{t-C_4H_9CN}$. For the shorter collapsing time, we always get a lower abundance of t, the same as that obtained by avoiding reaction 22 with set 3.

\cite{garr17} derived an upper limit of $0.59$ for the peak abundance ratio between $\rm{n-C_4H_9CN}$ and $\rm{n-C_3H_7CN}$.
Fig. \ref{fig:but_pc-rat} shows this ratio
with the warm-up time scale (left) and collapsing time scale (right). The upper limit is shown with the red horizontal line. Modelled peak ratio is less than the derived upper limit for most of the cases, except for set 2 and shorter collapsing time ($\rm{t_{coll}<5 \times 10^5}$ years).

It was interesting to see whether the observed ratio between the i and n $\rm{C_3H_7CN}$ is also sustained between the i and n of $\rm{C_4H_9CN}$ or gets amplified or reduced. Fig. \ref{fig:in-butyl} shows the i/n ratio of $\rm{C_4H_9CN}$ for different warm-up (left) and collapsing time (right) scale. 
For set 1, the i/n ratio for ${\rm C_3H_7CN}$ varies in the range $0.8-1.4$, whereas for $\rm{C_4H_9CN}$, it is $0.69-2.37$. With the same BE value, various warming times, and two types of activation barriers, \cite{garr17} obtained an i/n ratio in the range of $0.6-2.2$, which is in excellent agreement with our model. With the set 2 and 3 BEs, the i/n ratio of $\rm{C_4H_9CN}$ is significantly changed. Comparing Fig. \ref{fig:in-butyl} (i/n ratio of $\rm{C_3H_7CN}$) and Fig. \ref{fig:PC_rat} (i/n ratio of ${\rm C_4H_9CN}$), it is clear that the branching is more favourable for the higher-order alkyl cyanides with the realistic BE sets.

\cite{garr17} obtained a s/n, t/n, and (i+s+t)/n ratio of $\rm{C_4H_9CN}$ in the range $1.7-4.3$, $0.015-0.1$, and $3.0-5.8$, respectively. With the set 1, we obtain these ratios $2.23-3.03$, $0.19-1.52$, and $3.2-6.38$, respectively (see Table \ref{tab:ratios}). All these ratios obtained with the set 2 and set 3 BEs are also noted in Table \ref{tab:ratios} and compared with previous observations and modeling results.

In general, there are no systematic differences between the results predicted by our models and \cite{garr17}. However, with our more realistic BE values, set 3 \citep[low BEs compared to that used in][]{garr17}, we find that the branching is more favourable. Furthermore, a major difference is obtained when one hydrogen abstraction reaction (reaction 22) of $\rm{i-C_3H_7CN}$ is included. It yields a high abundance of $\rm{t-C_4H_9CN}$. Due to the exceptional environments of the galactic centre region, none of our models and models of \cite{garr17} could possibly explain the observed high abundance of some of the BCMs \citep{bonf19,will20}. 
The BEs used by \cite{garr17} were some educated estimations. So, we would refer to using our set 3 BEs obtained with the tetramer water substrate for future modeling. We notice a huge impact of collapsing (Model A) and warm-up (Model B) time scales in the abundances of these species. Based on the obtained results, we recommend using relatively moderate collapsing and warm-up time scales ($t_{\rm coll}$=$t_{\rm warm1} \sim 3-5 \times 10^5$ years).

\section{Conclusions}
\label{sec:conclusions}

We carry out an extensive study on the formation of various BCMs. Some of our significant findings from this work are:

\begin{itemize}
 \item One of the critical parameters for astrochemical modeling is the BE of species. Here, we provide a realistic estimation of the BEs for some BCM-related species for the first time. Noticeably lower BE values \citep[as compared to previously used][]{garr17} for $\rm{CH_2CHCN}$ (3540 K), $\rm{C_2H_5CN}$ (4886 K), $\rm{C_3H_7CN}$ (5567 K, 4970 K for n and i $\rm{C_3H_7CN}$, respectively), and $\rm{C_4H_9CN}$ (4388 K, 5151 K, 5313 K, and 5148 K for n, i, s, and $\rm{t-C_4H_9CN}$, respectively) are obtained. \\
  
 \item The enthalpies of formation, polarizabilities, dipole moments, and activation barriers through TS calculations are calculated quantum chemically to better estimate the modeling results. \\
 
\item \cite{bell14} observed i-PC/n-PC ratio of $\sim 0.4 \pm 0.06$.
With set 3, we obtain i-PC/n-PC $\sim$ 0.7 when a shorter warmup ($t_{warm1}=10^5$ years for Model A) and shorter collapsing time ($t_{coll}=1 - 3 \times 10^5$ years for Model B) are used.
With the set 2 and set 3, it is observed that the abundances of these species are greatly affected. Compared to the modeled i/n ratio of $\rm{C_3H_7CN}$ (varies in the range $0.7-3.4$ for set 2 and 3 noted in Table \ref{tab:ratios}), the i/n ratio of $\rm{C_4H_9CN}$ is greatly enhanced ($0.12-67$ shown in Table \ref{tab:ratios}). Thus, the branching is more favourable for the higher-order alkyl cyanides. It is also noticed that with the increase in the warm-up and collapsing time, in general, for set 3 (consisting of most updated BEs) ratio of the i/n increases. \\

\item Here, for the destruction of $\rm{i-C_3H_7CN}$, we propose one hydrogen abstraction reaction (reaction 22). The TS calculation of this reaction yields an activation barrier of $947$ K. Inclusion of this reaction drastically increases the abundance of $\rm{t-C_4H_9CN}$ by the CH$_3$ addition of $\rm{CH_3(C)CH_3CN}$ (reaction 36). Furthermore, we found that the formation of $\rm{t-C_4H_9CN}$ is favourable when a longer collapsing time is used along with reaction 22.

\end{itemize}

\section*{Acknowledgements}
S.S. acknowledges Banaras Hindu University and UGC, New Delhi, India, for providing a fellowship. M.S. would like to acknowledge the financial support from S. N. Bose National Centre for Basic Sciences, Salt Lake, Kolkata under the Department of Science and Technology (DST), Government of India. P.G. acknowledges support from a Chalmers Cosmic Origins postdoctoral fellowship. A.P. acknowledges financial support from the IoE grant of Banaras Hindu University (R/Dev/D/IoE/Incentive/2021-22/32439) and financial support through the Core Research Grant of SERB, New Delhi (CRG/2021/000907). AD would like to acknowledge ICSP for support.

\section*{Data Availability}

The data underlying this article will be shared on reasonable request to the corresponding author.



\bibliographystyle{mnras}
\bibliography{BCM_mnras.bbl}

\clearpage

\appendix

\section{Potential energy surfaces}
\begin{figure*}
\centering
\includegraphics[width=\columnwidth,keepaspectratio]{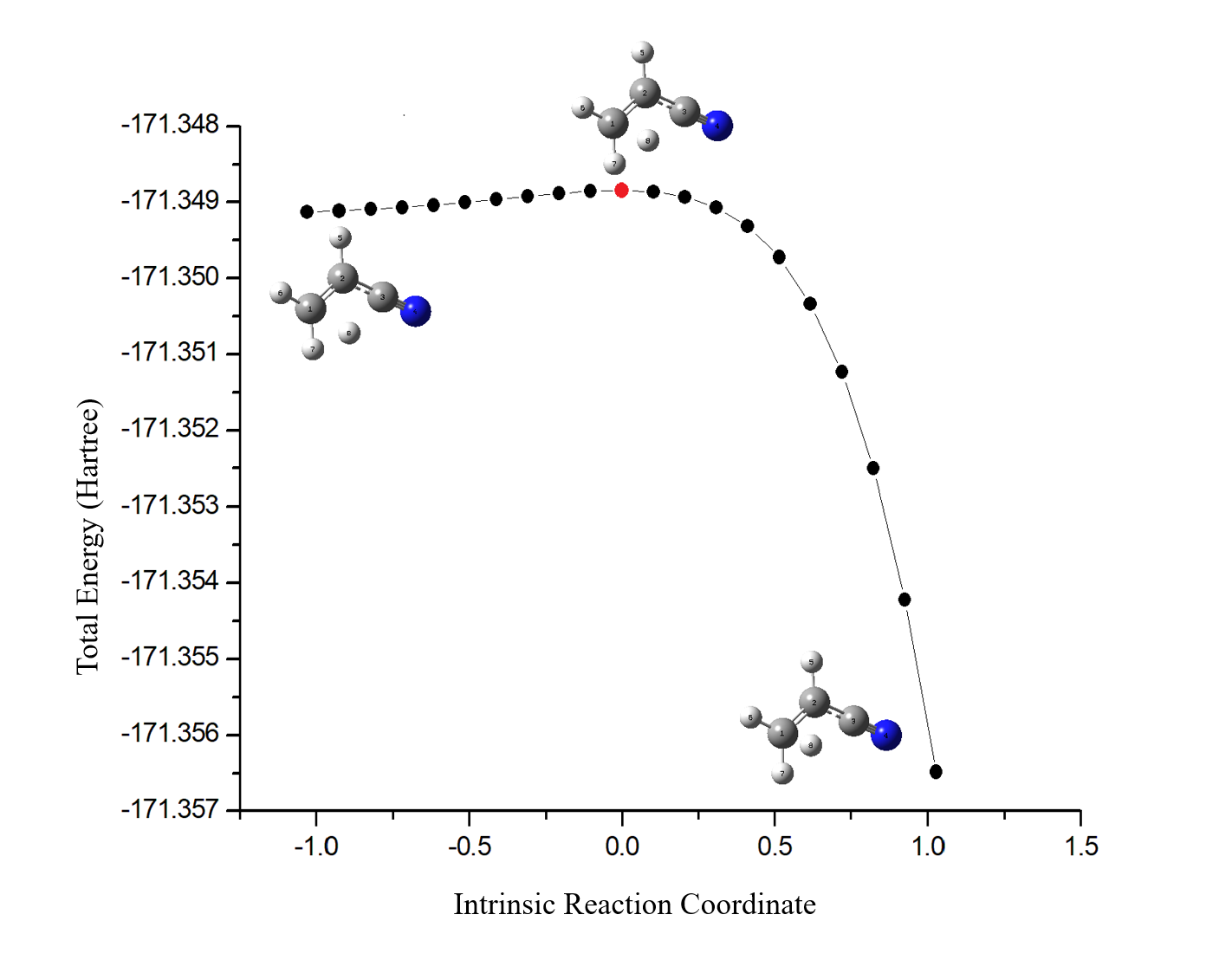}
\includegraphics[width=\columnwidth,keepaspectratio]{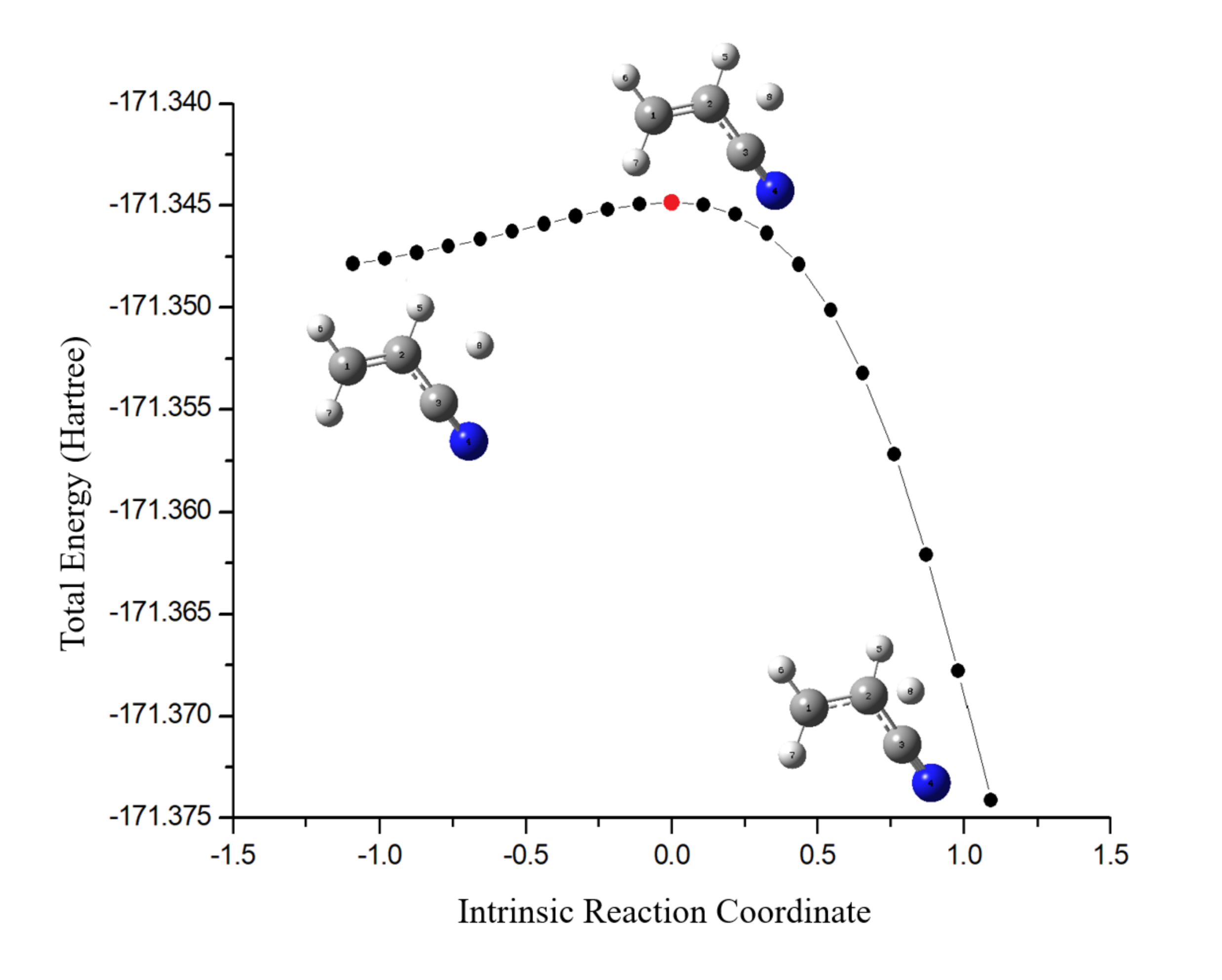}
\caption{Potential energy surfaces for reactions 5 and 6.
\label{fig:reaction_5}}
\end{figure*}

\begin{figure}
\centering
\includegraphics[width=\columnwidth,keepaspectratio]{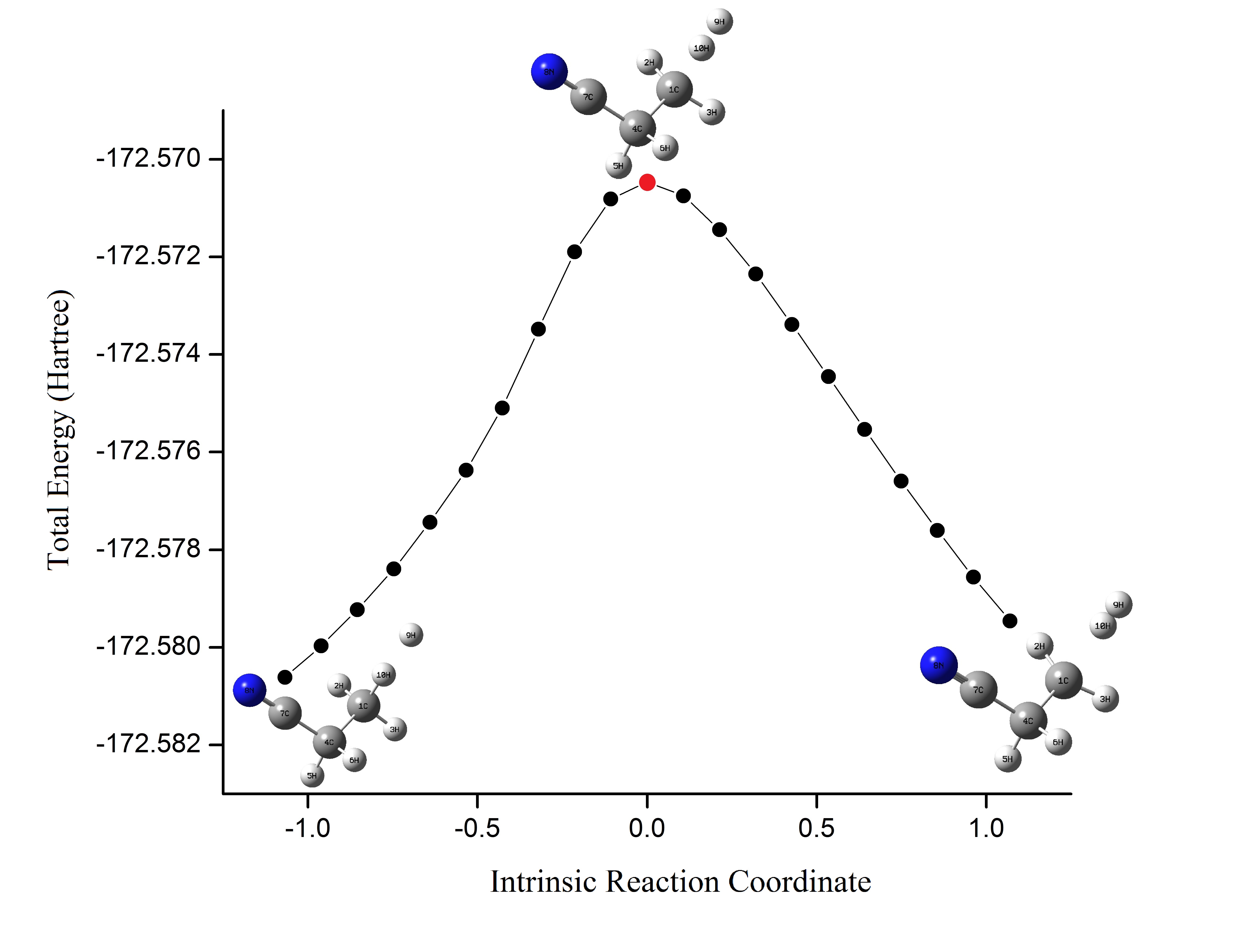}
\caption{Potential energy surface for reaction 11.
\label{fig:ch2ch2cn}}
\end{figure}

\begin{figure}
\centering
\includegraphics[width=\columnwidth,keepaspectratio]{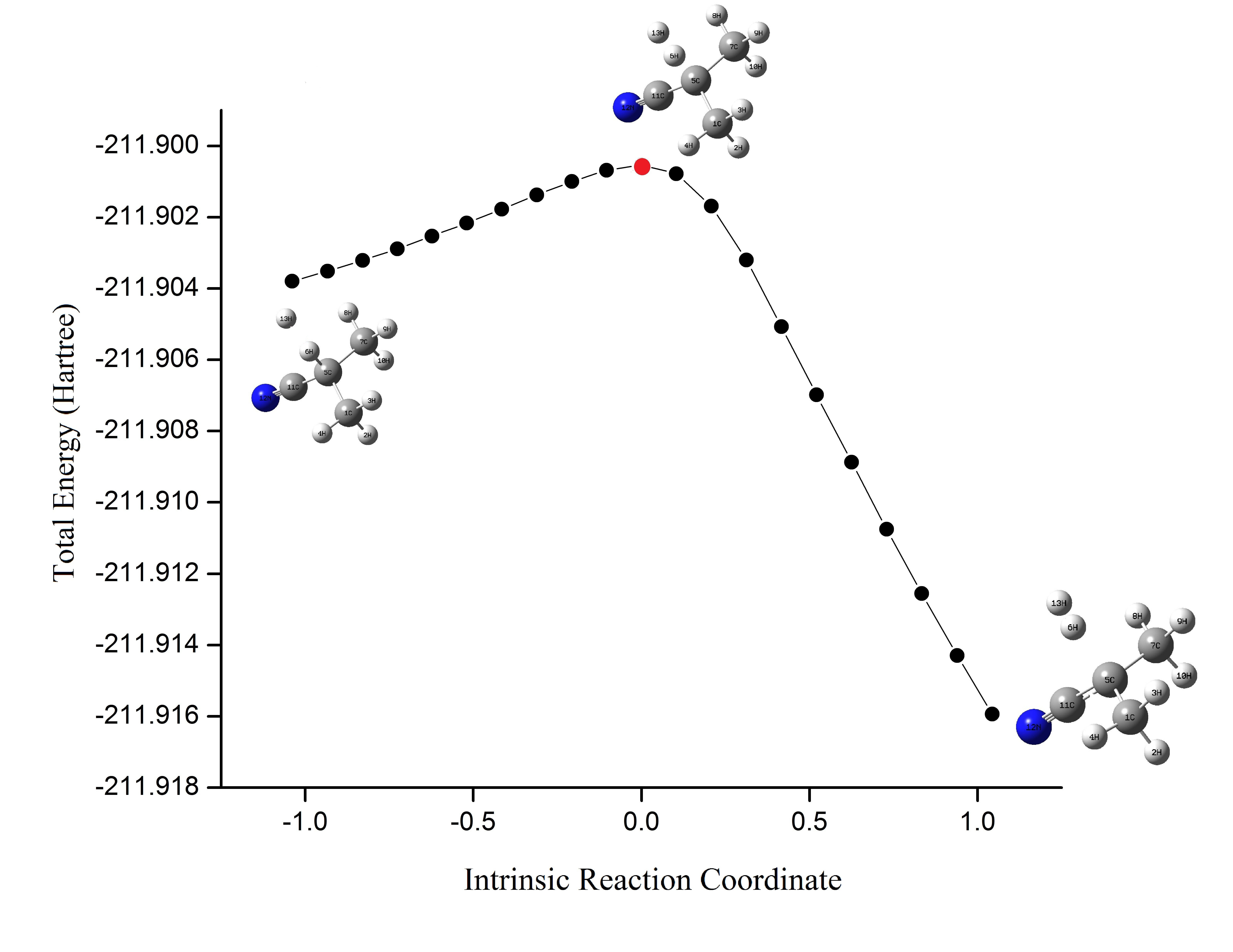}
\caption{Potential energy surface for reaction 22.
\label{fig:ch3cch3cn}}
\end{figure}

Potential energy surface diagrams of the ice-phase chemical reactions 5, 6, 11, and 22 noted in Tables \ref{tab:EV} and \ref{tab:prop} are shown in Figs. \ref{fig:reaction_5}, \ref{fig:ch2ch2cn}, and \ref{fig:ch3cch3cn}, respectively.

\clearpage

\section{Polarizability and dipole moment}
We quantum chemically calculate the polarizability and total dipole moment of BCM-related species noted in Table \ref{tab:polar_dipole} using the Gaussian 09 suite.
For these calculations, we use the DFT-B3LYP/6-31G(d,p) level of theory. Our calculated values are compared with the existing experimental values (if available in the NIST WebBook database).

\begin{table*}
{\centering
  \caption{Calculated polarizability and total dipole moment of some BCM-related species.}
  \label{tab:polar_dipole}
  \begin{tabular}{cccccc}
    \hline
Serial &   & \multicolumn{2}{c}{Polarizability (\AA$^3$)} & \multicolumn{2}{c}{Total dipole moment (D)} \\
   \cline{3-6}
 &  Species & This Work & Experimental$^a$ & This Work & Experimental$^b$ \\ 
 &  & DFT-B3LYP 6-31G(d,p) & & DFT-B3LYP 6-31G(d,p) & \\ 
   \hline
  1 &      $\rm{CH_3CN}$ & 3.4378 & 4.280 & 3.8279 & 3.92 \\
  2 &      $\rm{C_2H_2CN}$ & 4.6441 & --- & 3.2203 & --- \\
  3 &      $\rm{CH_2CHCN}$ & 4.9597 & --- & 3.8772 & 3.87 \\
  4 &      $\rm{\dot CH_2CH_2CN}$ & 4.7300 & --- & 3.6998 & --- \\
  5 &      $\rm{CH_3 \dot CHCN}$ & 5.1242 & --- & 3.8985 & --- \\
  6 &      $\rm{C_2H_5CN}$ & 5.0442 & 6.240 & 3.9198 & 4.02 \\
  7 &      $\rm{\dot CH_2CH_2CH_2CN}$ & 6.3156 & --- & 3.8816 & --- \\
  8 &      $\rm{CH_3\dot CHCH_2CN}$ & 6.4000 & --- & 3.7363 & --- \\
  9 &      $\rm{CH_3CH_2\dot CHCN}$ & 6.8268 & --- & 4.0005 & --- \\
 10 &      $\rm{n-C_3H_7CN}$ & 6.6846 & 8.400 & 4.0636 & 4.07 \\
 11 &      $\rm{\dot CH_2CH(CH_3)CN}$ & 6.3615 & --- & 3.9399 & --- \\
 12 &      $\rm{\dot CH_3\dot C(CH_3)CN}$ & 6.8268 & --- & 4.1409 & --- \\
 13 &      $\rm{i-C_3H_7CN}$ & 6.6327 & 8.049 & 3.9515 & --- \\
 14 &      $\rm{CH_3CH_2\dot CHCH_2CN}$ & 8.0019 & --- & 3.8328 & --- \\
 15 &      $\rm{n-C_4H_9CN}$ & 8.3264 & --- & 4.1606 & 4.12 \\
 16 &      $\rm{i-C_4H_9CN}$ & 8.2272 & --- & 3.9850 & --- \\
 17 &      $\rm{s-C_4H_9CN}$ & 8.2123 & --- & 3.9019 & --- \\
 18 &      $\rm{t-C_4H_9CN}$ & 8.1946 & 9.591 & 3.9600 & 3.95 \\
 19 &      $\rm{\dot CH_2CH_2CH_3}$ & 4.7330 & --- & 0.2533 & --- \\
 20 &      $\rm{CH_3\dot CHCH_3}$ & 4.8263 & --- & 0.2042 & --- \\
 21 &      $\rm{C_3H_8}$ & 5.0471 & 5.921 & 0.0491 & 0.08 \\
 22 &      $\rm{\dot CH_2CH_2CH_2CH_3}$ & 6.3082 & --- & 0.2525 & --- \\
 23 &      $\rm{CH_3\dot CHCH_2CH_3}$ & 6.4519 & --- & 0.2163 & --- \\
 24 &      $\rm{n-C_4H_{10}}$ & 6.6446 & 8.020 & 0.0000 & 0.00 \\
 25 &      $\rm{\dot CH_2CH(CH_3)CH_3}$ & 6.333 & --- & 0.1989 & --- \\
 26 &      $\rm{CH_3\dot C(CH_3)CH_3}$ & 6.5201 & --- & 0.1990 & --- \\
 27 &      $\rm{i-C_4H_{10}}$ & 6.6105 & 8.009 & 0.0765 & 0.13 \\
 28 &      $\rm{n-C_5H_{12}}$ & 8.2598 & 9.879 & 0.0477 & --- \\
 29 &      $\rm{i-C_5H_{12}}$ & 8.1768 & 8.770 & 0.0542 & 0.13 \\
 30 &      $\rm{neo-C_5H_{12}}$ & 8.1427 & 10.240 & 0.0001 & 0.00 \\
        \hline
    \end{tabular}} \\
$^a$ \url{https://cccbdb.nist.gov/xp1.asp?prop=9} \\
$^b$ \url{https://cccbdb.nist.gov/xp1.asp?prop=7}    
\end{table*}






\bsp	
\label{lastpage}
\end{document}